%% file: continuous_energy_release.tex
\documentclass[useAMS,usenatbib]{mnras}
\usepackage{graphicx,natbib,amssymb,amsmath,stmaryrd,makecell}
\usepackage{txfonts}
\usepackage{xcolor}
\usepackage{hyperref}
\usepackage{xspace}

\input{Befehle}

\newcommand{\COBEF}{{\it COBE/FIRAS}\xspace}
\newcommand{\PIXIE}{{\it PIXIE}\xspace}

\newcommand{\DlnrhoCMB}{\left.\frac{\Delta\rho_\gamma}{\rho_\gamma}\right|_{\rm CMB}}
\newcommand{\DlnrhoCMBt}{\Delta\rho_\gamma/\rho_\gamma\big|_{\rm CMB}}

\newcommand{\Dlnrhoin}{\left.\frac{\Delta\rho_\gamma}{\rho_\gamma}\right|_{\rm in}}
\newcommand{\Dlnrhoint}{\Delta\rho_\gamma/\rho_\gamma\big|_{\rm in}}

\title[Continuous large energy release]{CMB spectral distortions from continuous large energy release}

\begin{document}

\author[Acharya et al.]
{Sandeep Kumar Acharya$^1$\thanks{E-mail:sandeep.acharya@manchester.ac.uk}
and 
Jens Chluba$^1$\thanks{E-mail:jens.chluba@manchester.ac.uk}
\\
$^1$Jodrell Bank Centre for Astrophysics, School of Physics and Astronomy, The University of Manchester, Manchester M13 9PL, U.K.
}

\date{\vspace{-0mm}{Accepted 2021 --. Received 2021 --}}

\maketitle

\begin{abstract}
Accurate computations of spectral distortions of the cosmic microwave background (CMB) are required for constraining energy release scenarios at redshifts $z\gtrsim 10^3$. The existing literature focuses on distortions that are small perturbations to the background blackbody spectrum. At high redshifts ($z\gtrsim 10^6$), this assumption can be violated, and the CMB spectrum can be significantly distorted at least during part of its cosmic evolution. 
In this paper, we carry out accurate thermalization computations, evolving the distorted CMB spectrum in a general, fully non-linear way, consistently accounting for the time-dependence of the injection process, modifications to the Hubble expansion rate and relativistic Compton scattering. Specifically, we study single energy injection and decaying particle scenarios, discussing constraints on these cases. 
We solve the thermalization problem using two independent numerical approaches that are now available in {\tt CosmoTherm} as dedicated setups for computing CMB spectral distortions in the large distortion regime.
New non-linear effects at low frequencies are furthermore highlighted, showing that these warrant a more rigorous study.
This work eliminates one of the long-standing simplifications in CMB spectral distortion computations, which also opens the way to more rigorous treatments of distortions induced by high-energy particle cascade, soft photon injection and in the vicinity of primordial black holes.
\end{abstract}


\section{Introduction}

Cosmic microwave background (CMB) spectral distortions provide a unique probe of standard and non-standard cosmological scenarios in the early Universe (redshift $z\gtrsim 10^3$). At these redshifts, the background electrons and CMB photons are tightly coupled and in close thermal equilibrium. Energy injection to the baryon-photon fluid can disturb this equilibrium, creating spectral distortions to the CMB blackbody distribution. 
The distorted CMB spectrum carries information of the epoch at which energy was injected to the background baryon-photon fluid. The magnitude of the spectral distortion depends on the amount 
 of energy that was injected at a given epoch and can be related to parameters of energy injection scenarios such as abundance of dark matter, primordial black holes, etc.  
Combined with the current constraints on the CMB spectral distortions obtained with \COBEF \citep{Fixsen1996}, we can obtain strong limits on the allowed parameter space for many scenarios \citep{Chluba2011therm, Sunyaev2013, Lucca2020}.  

To constrain the parameter space accurately, we need a detailed understanding of the microphysical interaction between the CMB photons and background electrons. At $z\lesssim 2\times 10^6$, Compton scattering is the dominant collision process relevant to the formation of energy release distortion signals in the CMB spectral band ($\nu\simeq 10-10^3 \, \GHz$). 
During the Comptonization era, several regimes can be distinguished.
At $z\lesssim 10^4$, Compton scattering is inefficient and background electrons at a higher temperature (due to energy injection) can boost the CMB photons to create a $y$-distortion \citep{Zeldovich1969}. The Comptonization process becomes increasingly efficient at higher redshifts, driving the distorted CMB spectrum towards a Bose-Einstein distribution \citep[or $\mu$-type distortion, see][]{Sunyaev1970mu}.
The long-standing constraints on $y$ and $\mu$-distortions from \cite{Fixsen1996} are $1.5\times 10^{-5}$ and $9\times 10^{-5}$, respectively. In terms of energy release to the CMB photon field this implies $\Delta \rho_\gamma/\rho_\gamma \lesssim \pot{6}{-5}$ ($95\%$~c.l.), meaning that large energy release in the Comptonization era is excluded.

At intermediate redshifts ($10^4\lesssim z \lesssim \pot{2}{5}$), the distortions created by Comptonization can furthermore show a richer phenomenology \citep{Illarionov1974, Hu1995PhD, Chluba2011therm, Khatri2012mix, Chluba2013Green}, with a signal that is not just given by a simple sum of $y$ or $\mu$ distortions. Additional spectral diversity can be created by photon injection processes \citep{Chluba2015GreensII, Bolliet2020PI} and non-thermal electron populations found in high-energy particle cascades \citep{Ensslin2000, Slatyer2015, Acharya2019a}.
These refinements could allow us to distinguish various sources of distortions using future CMB spectroscopy.

However, Compton scattering alone, being a photon number conserving process, does not allow the distorted CMB spectrum to relax back to  a Planckian spectrum. 
At $z\gtrsim 2\times 10^6$, photon number non-conserving processes such as Bremsstrahlung and Double Compton become efficient  \citep{Sunyaev1970mu,Danese1982,Burigana1991}. These emission processes, combined with Compton scattering, tend to wash out distortions that are already created. The thermalization efficiency can be captured using the distortion visibility function, which is just defined as the ratio of energy density of the surviving distortion today and the energy density that was injected into the photons. While the distortion visibility function is close to unity at $z\lesssim 10^5$ \citep[e.g.,][]{Chluba2014}, it decays exponentially at increasing redshifts. Therefore, more and more energy can generally be injected to the baryon-photon fluid without violating the current spectral distortion bounds.

Almost all of the previous literature assumes the distortion to be small for the computation of the constraints on various energy injection scenarios. The "smallness" of the distortion essentially requires the problem to be {\it linear}. This means that the distorted spectrum between different injection scenarios can be related to each other just by rescaling each spectrum. For example, if we have the distorted spectrum for a decaying particle with abundance $f_{\rm dm}$ (w.r.t total dark matter), we can obtain the spectrum for the same with 10 times higher or lower abundance by just scaling  the given spectrum up or down by a factor of 10. Therefore, for small distortions we can create a library of solutions by running the thermalization code {\tt CosmoTherm} \citep{Chluba2011therm} just once and then rescale these solutions appropriately. 

\cite{Chluba2020large} (henceforth CRA20) went beyond the small distortion limit and showed that the criteria of linearity can be violated at $z\gtrsim \pot{5}{6}$ when the distortions can be of the order unity relative to the CMB blackbody spectrum, at least for part of the evolution. This makes the thermalization problem {\it non-linear}, i.e., the distortion spectrum becomes a function of magnitude of the distortion already present and cannot be just scaled to obtain another solution. %
This complicates matters as one has to solve the thermalization problem for every combination of energy injection parameters. 
In addition, relativistic corrections to the Compton process have to be considered using a scattering kernel treatment based on {\tt CSpack} \citep{CSpack2019}, rendering one computation time-consuming.

CRA20 demonstrated that the solution space is much richer and that the distortion visibility function can differ drastically from the one obtained in the small distortion limit. However, the authors assumed that the distorted CMB spectrum evolves along a sequence of quasi-stationary solutions, which is a good approximation at high redshifts since the scattering processes are much faster than the Hubble rate, but breaks down at later times. The authors furthermore only studied single energy injection scenarios (i.e., a $\delta$-function in redshift), which cannot be easily generalized to injection scenario from decaying particles without taking into account the full time-dependence of the problem.

In this paper, we revisit the calculation of CRA20 and perform a more rigorous calculation, embedding the full non-linear and time-dependent thermalization treatment into {\tt CosmoTherm}. We evolve the photon spectrum without assuming quasi-stationarity or linearity, implementing it in two independent ways to study possible physical/numerical limitations of the used approaches. 
First, we follow the standard approach, which solves a parabolic partial differential equation \citep[i.e., the Kompaneets equation][]{Kompa56} for photon evolution and energy exchange with the background electrons, but here in a fully non-linear manner. The Kompaneets equation assumes that Compton scattering between the photons and electrons remains non-relativistic, an approximation that can be violated at $z\simeq 10^7$ and at high photon energies, hence suggesting difference with the exact solution to become visible \citep{Acharya2021FP}. 

In the second approach, we remove the assumptions underlying the Kompaneets equation by solving the exact Boltzmann equation (in the isotropic limit) for the photons using the relativistic Compton scattering kernel of {\tt CSpack}. While there are interesting physical differences between the solutions obtained using the aforementioned approaches \citep[see also][]{Acharya2021FP}, here we do not see significant differences at the level of the constraints on parameters of energy injection scenarios. We therefore pre-dominantly use the first approach, which is less time-consuming. This is because in the scattering kernel approach the electron temperature is evolving with the photon spectrum such that the kernel has to be computed at the corresponding electron temperature, making it numerically expensive. 

To highlight the importance of non-linear terms, we compare the exact spectrum with the one obtained in small distortion approximation and show that the approximate solution becomes unphysical for large energy injection. We furthermore explicitly compute the distortion visibility function for single energy injection scenarios and decaying particles. We derive constraints on allowed parameters for both cases and show that these are vastly different from small distortion approximations for single energy injection. However, the constraints for decaying particles remains similar to the small distortions case, as energy is injected in a broad redshift range, which does not allow the non-linearity of the problem to take hold. We finish our study by considering line photon injection cases \citep{Chluba2015GreensII,Bolliet2020PI} for which at $z\lesssim 10^4$ non-linear effects due to stimulated Compton scattering can have interesting consequences.

The paper is structured as follows. In Sec. \ref{sec:thermalization}, we describe the non-linear terms and modifications to {\tt CosmoTherm} in some detail. We also briefly describe the evolution equation for the photons and the temperature of the background electrons. We describe the qualitative differences between non-linear and linear spectral distortion solutions in Sec. \ref{sec:soln_comp}. We show our main results in terms of constraints for single energy cases and decaying particle scenario in Sec. \ref{sec:single_energy_case} and \ref{sec:decay}, respectively. In Sec. \ref{sec:low_freq_photon}, we study a couple of cases of low-frequency photon injection and describe the qualitative differences w.r.t. the linear treatment. We then conclude in Sec.~\ref{sec:conclusion}. 

\section{Treatment of the thermalization physics}

\label{sec:thermalization}
In this section, we provide several details about how the treatment of the cosmological thermalization problem has to be augmented in the regime of large distortions. The starting point is {\tt CosmoTherm} \citep{Chluba2011therm}, which we extend here to include non-linear terms in the spectral distortion. We furthermore add an exact kernel treatment for Compton scattering.

The photon Boltzmann equation together with the coupled electron temperature equation can be formally expressed as
\bsub
\label{eq:Boltzmann_system}
\begin{align}
\label{eq:Boltzmann_system_a}
\frac{\id n}{\id\tau}
&=
 \left.\frac{\id n}{\id\tau}\right|_{\rm CS}
+\left.\frac{\id n}{\id\tau}\right|_{\rm e/a}
+\left.\frac{\id n}{\id\tau}\right|_{\rm S},
\\
\label{eq:Boltzmann_system_b}
\frac{\id \rho}{\id\tau}&=
 \left.\frac{\id \rho}{\id\tau}\right|_{\rm CS}
+\left.\frac{\id \rho}{\id\tau}\right|_{\rm e/a}
+\left.\frac{\id \rho}{\id\tau}\right|_{\rm h}
+\left.\frac{\id \rho}{\id\tau}\right|_{\rm exp}.
\end{align}
\esub
Here, $n$ is the isotropic photon occupation number, $\rho=\The/\Thz$ where $\The=k \Te/\me c^2$ and $\Thz=k \Tz/\me c^2$ are the dimensionless electron and reference blackbody temperature.
We will explain each of the terms below, but the various subscripts are 'CS' for Compton scattering terms, 'e/a' for BR and DC emission and absorption terms, 'S' for extra photon sources, 'h' for external heating terms, and 'exp' for the Hubble expansion.

In Eq.~\eqref{eq:Boltzmann_system}, the overall time-coordinate is the Thomson optical depth, $\tau=\int \Ne \sigT c \id t$. The solution for the free electron number density, $\Ne$, is usually obtained using {\tt CosmoRec} \citep{Chluba2010b}, but can also be explicitly calculated within {\tt CosmoTherm} \citep[see][for details]{Bolliet2020PI}.
Coulomb collision are assumed to be very efficient and keep the electron distribution function close to a Maxwellian at all times. This assumption can be violated at late times and also when high energy particle cascades are involved \citep[e.g.,][]{Slatyer2015, Acharya2019a}.

The isotropic photon occupation number is expressed in terms of $n(x, \tau)$, where $x=h\nu/k\Tz$, where $\Tz$ is the CMB blackbody reference temperature. Importantly, $T_z$ is chosen to explicitly scale as $T_z\propto a^{-1}= (1+z)$ to absorb the effect of redshifting on the average spectrum, which simplifies the numerical treatment immensely \citep[see][for discussion]{Chluba2011therm}. In the above, $a$ is the cosmological scale factor.

We now explain the different collision terms and how they are augmented in comparison to the previous treatment in {\tt CosmoTherm}. Most importantly, we will no longer assume that the spectral distortion remains small.
The solutions of Eq.~\eqref{eq:Boltzmann_system} can then be obtained numerically using the solvers of {\tt CosmoTherm}. Details about how to discretize the photon evolution equation are also be given below. 
We start with the external photon source and heating terms.

\subsection{External photon source and heating term}
In this work, we consider three scenarios i) single energy release ($\delta$-function in $z$), ii) energy injection from dark matter decay and iii) single photon injection. The released energy in cases i) and ii) is assumed to be dissipated as heat (i.e., pure energy release), which always is good approximation at $z\gtrsim 10^5$. In {\tt CosmoTherm}, this is treated as a heating term in the electron temperature equation. Through Compton scattering, this then causes the upscattering of photons and formation of distortions. No photon source term is added directly for i) and ii). We model the single energy release as a narrow Gaussian profile in redshift with the width of Gaussian being a couple of percent of injection redshift.

For case ii). the matter energy injection rate from dark matter decay of a particle with rest mass $M_X$ can be written as
\begin{equation}
%
\dot{Q}_X=-\frac{\id a^3 \rho_X}{a^3 \id t}=M_X c^2\,f_X N_{\rm{H}}\Gamma_X \expf{-\Gamma_X\,t},   
\label{eq:decay_source}
\end{equation}
where $f_X$ parameterizes the initial amount of dark matter relative to the number density of hydrogen nuclei, $N_{\rm{H}}$, and $\Gamma_X$ is the inverse lifetime of the particle. For $t\ll \Gamma_X^{-1}$ (timescale much smaller than lifetime of decay), the energy density of dark matter simply scales as $a^{-3}$. For $t\gtrsim \Gamma_X^{-1}$, we have exponential depletion of the dark matter number and energy density, with the energy being transferred to the kinetic energy of the baryons. 

For single photon injection at $z_{\rm in}$, we approximate the photon spectrum as a narrow Gaussian to mimic a source term
\begin{equation}
\left.\frac{\id n(x)}{\id \tau}\right|_{\rm S}= A\frac{\delta (x-x_{\rm in})}{x^2}.
\end{equation}
Here, the amplitude $A$ is determined by both the injected photo number, $\Delta N_\gamma/N_\gamma$, and the injection frequency $x_{\rm in}=h \nu_{\rm in}/k\Tz$.

For photon injection scenarios, not only the total injected energy but also the number of injected photons determine the distortion. In this case, a negative $\mu$ distortions can arise at high redshifts ($z\gtrsim \pot{2}{5}$), while at low redshifts ($z\lesssim \pot{2}{5}$) a rich spectral signal can be formed \citep[see][for details and examples]{Chluba2015GreensII,Bolliet2020PI}. This case is used to illustrate the importance of non-linear scattering terms at low-frequencies, which have been neglected in the thermalization problem so far.

\subsection{Compton scattering terms}
Out of all the terms in Eq.~\eqref{eq:Boltzmann_system}, those related to the Compton process are most complicated. Here we provide two independent formulations and explain how these are discretized inside {\tt CosmoTherm}. The corresponding electron temperature equation contributions will be discussed in Sect.~\ref{sec:Te-equation}

\subsubsection{Kompaneets terms with large distortions}
The standard diffusion approximation for the Compton collision term is given by the  Kompaneets equation \citep{Kompa56, Burigana1991, Chluba2011therm}
\begin{equation}
\left.\frac{\id n}{\id\tau}\right|_{\rm CS}
=\frac{\The}{x^2}\frac{\partial }{\partial x}x^4\left[\frac{\partial n}{\partial x}+\phi\, n(n+1)\right],
\label{eq:Kompaneets_eq}
\end{equation}
where we introduced $\phi=T_z/\Te$ for convenience. Using $n=n_{\rm{pl}}+\Delta n$, Eq.~\eqref{eq:Kompaneets_eq} is can be cast into the form
\begin{align}
\left.\frac{\id \Delta n}{\id\tau}\right|_{\rm CS}
&\!=\!\frac{\The}{x^2}\frac{\partial }{\partial  x}x^4\left[\frac{\partial \Delta n}{\partial x}+\Delta\phi n_{\rm{pl}}(n_{\rm{pl}}+1)+\phi(1+2n_{\rm{pl}} + \Delta n)\Delta n \right],
\nonumber
\end{align}
where $\Delta\phi=\phi-1$. To arrive at above expression, we have used the fact that $n_{\rm{pl}}=1/(\expf{x}-1)$ is the reference blackbody spectrum, which satisfies the equation $\id n_{\rm{pl}}/\id\tau=0$ for $x=h\nu/k T$ when $T\propto (1+z)$ is chosen. We furthermore used $\partial_x n_{\rm{pl}}=-n_{\rm{pl}}(1+n_{\rm{pl}})$.
For computational reasons it is good to explicitly carry out all the frequency derivatives and write the Kompaneets equation as a combination of terms $\propto \Delta n, \partial_x \Delta n$ and $\partial^2_x \Delta n$.
After a few simplifications, we find
\begin{align}
\label{eq:Kompaneets_distortion_simpl}
\left.\frac{\id \Delta n}{\id\tau}\right|_{\rm CS}
&=D_{\rm e}\frac{\partial^2\Delta n}{\partial x^2}+D_{\rm e}\left[\frac{4}{x}+\phi \xi \right]\frac{\partial \Delta n}{\partial x}+D_{\rm e}\phi \xi \left[\frac{4}{x}+
    \frac{\partial \ln\xi}{\partial x}\right]\Delta n 
\nonumber
\\ 
&\quad-D_{\rm e}\Delta\phi\zeta\left[\frac{4}{x}-\xi\right]+2D_{\rm e}\phi \Delta n\frac{\partial \Delta n}{\partial x}+D_{\rm e}\phi\left(\frac{4}{x}\right) \Delta n^2,
\end{align}
where $D_{\rm e}=\The x^2$, $\xi=1+2n_{\rm{pl}}\equiv \coth(x/2)$ and $\zeta=-n_{\rm{pl}}(n_{\rm{pl}}+1)$. Equation~\eqref{eq:Kompaneets_distortion_simpl} is exactly same as Eq.~(7) of \cite{Chluba2011therm} except for the last two term, which arise from quadratic contributions in $\Delta n$ that previously were neglected. These describe enhanced stimulated recoil drifts, which are usually most relevant at $x\ll 1$.

From Eq.~\eqref{eq:Kompaneets_distortion_simpl}, we can also see that in the absence of distortions, distortions are sourced by the term $-D_{\rm e}\Delta\phi\zeta\left[\frac{4}{x}-\xi\right]\equiv Y(x)[\The-\Thz]$ if $\The\neq \Thz$. Here, $Y(x)=G(x)\left[x\coth(x/2)-4\right]$ is the photon occupation number of a $y$-type distortion and $G(x)=x\expf{x}/(\expf{x}-1)^2$ that of a small temperature shift. Signals from heating thus always start out as $y$-type distortions.

\subsubsection{Discretizing the photon spectrum and its derivatives}
To numerically treat the photon diffusion problem, Eq.~\eqref{eq:Kompaneets_distortion_simpl} has to be discretized. This is achieved by expressing $\Delta n$ is terms of Lagrange polynomials, $L_j^{(i)}(x)$, which are defined for a given set of points around the considered frequency grid point $i$. For instance, well inside the computational domain, we can use a fifth order Lagrange polynomial representation, with grid points $x_{i-2}, x_{i-1}, x_{i}, x_{i+1}$ and $x_{i+2}$ (i.e., an interval centered on $i$). Then the expression for $\Delta n(x)$ and the relevant Lagrange polynomials take the form 
\bsub
\begin{align}
\Delta n(x)&=\sum_{j=i-2}^{i+2}\,L^{(i)}_j(x)\Delta n_j 
\\
L^{(i)}_j(x)
&=\prod_{\substack{m=i-2\\ m\neq j}}^{i+2} 
\frac{x-x_m}{x_j-x_m},
\end{align}
\esub
with $\Delta n_j=\Delta n(x_j)$. Boundary terms can be treated by shifting indices.
In a similar manner, one can define the frequency derivatives of $\Delta n(x)$, using derivatives of $L_j^{(i)}(x)$. 
For grids with constant $\Delta x$, this leads to the standard $n$-point stencils for numerical derivatives\footnote{For example, $\left.\partial \Delta n/\partial x\right|_{x=x_i}\approx (n_{i+1}-n_{i-1})/2\Delta x$ for the first derivative and $\left.\partial^2 \Delta n/\partial x^2\right|_{x=x_i}\approx (n_{i+1}-2n_{i}+n_{i-1})/\Delta x^2$ 
for the second derivative when
using a centered 3-point stencil.}; however, with the Lagrange polynomials one can accommodate non-uniform grids.
By grouping derivative terms, this leads to a banded matrix formulation of Eq.~\eqref{eq:Kompaneets_distortion_simpl} for all linear terms in $\Delta n$. The non-linear terms can be added in a similar way.
Overall, the discretized Kompaneets equation can then be cast into the form
\begin{align}
\label{eq:Kompaneets_distortion_simpl_ij}
\left.\frac{\id \Delta n_i}{\id\tau}\right|_{\rm CS}
&=\sum_j \left[S^{\rm K}_{ij}(\Thz, \The) + \Delta n_i T^{\rm K}_{ij}(\Thz) \right]\Delta n_j
+(\The-\Thz) Y_i,
\end{align}
where $Y_i=Y(x_i)$. 
To give expressions for the scattering matrices, $S^{\rm K}_{ij}$ and $T^{\rm K}_{ij}$, we define
\bsub
\begin{align}
\frac{\partial \Delta n_i}{\partial x_i}
&=\sum_j D^{(1)}_{ij} \Delta n_j
\\
\frac{\partial^2\Delta n_i}{\partial x_i^2}
&=\sum_j D^{(2)}_{ij} \Delta n_j,
\end{align}
\esub
where the banded matrices $D^{(k)}_{ij}$ directly follow from the Lagrange polynomial discretization. This then yields
\bsub
\begin{align}
S^{\rm K}_{ij}
&=\The x_i^2
\left[
D^{(2)}_{ij}
+\left(\frac{4}{x_i}+\phi \xi_i\right)D^{(1)}_{ij}
+\phi \xi_i \left(\frac{4}{x_i}+
    \frac{\partial \ln\xi_i}{\partial x_i}\right)\delta_{ij}
\right]
\\
T^{\rm K}_{ij}
&=\Thz x_i^2 
\left[2D^{(1)}_{ij}
+
\frac{4}{x_i}\delta_{ij}
\right],
\end{align}
\esub
where $T^{\rm K}_{ij}=0$, if non-linear terms in the photon spectrum are to be neglected.
The non-linear terms can be added perturbatively using the ordinary differential equation (ODE) solver developed in \cite{Chluba2010a} for {\tt CosmoRec} \citep{Chluba2010b}.

\subsubsection{Integrals over the photon distribution}
\label{subsec:photon_distribution}
To compute energy exchange terms and also for the scattering kernel approach (see Sect.~\ref{sec:kernel_treat}), integrals over the photon distribution have to be carried out. The basic problem is to write
\begin{align}
\label{eq:weights}
\int f(x) \id x
&\approx \sum_i w_i f_i,
\end{align}
with appropriate weights $w_i$ for each frequency grid point. This can be further generalized to include weight functions $g(x)$ that ensure that certain integrals become exact. For example, when one is interested in the photon number integral $\Delta G_{2}=\int x^2 \Delta n(x) \id x$, one would choose $g(x)=x^2$ such that $\Delta G_{2}\approx \sum w^*_i g(x_i) \Delta n_i=\sum w^*_i x^2_i \Delta n_i$. 

Assuming a centered 5-point stencil for the Lagrange polynomials, we can then define generalized weights as,
\begin{align}
\label{eq:weights_definition}
w^*_i=
\int_{x_i^-}^{x_i^+} \frac{g(x')}{g(x_i)}
\left[\sum_{k=i-2}^{i+2} L^{(k)}_i(x')\right] \id x'
\end{align}
with $x_i^\pm=(x_i+x_{i\pm1})/2$ defining the frequency bin around $x_i$. In {\tt CosmoTherm}, we have the option to choose $g(x)=1$, $g(x)=x^2$ or $g(x)=x^3$, with all cases providing highly accurate results.

\vspace{-2mm}
\subsubsection{Non-linear scattering kernel treatment}
\label{sec:kernel_treat}
The Kompaneets equation is only valid when the involved electron and photon energies remain small \citep[see][for more rigorous discussion]{Acharya2021FP}. The exact Compton collision term is given by
\begin{align}
\left.\frac{\id  n}{\id\tau}\right|_{\rm CS}
=\int P(x\rightarrow x',\The)\left[\expf{\xe'-\xe}n'(1+n)-n(1+n')\right]\id x',
\end{align}
where $\xe=h\nu/k\Te$ and $P(x\rightarrow x',\The)$ is the Compton scattering kernel which can be computed using {\tt CSpack}. 
It is convenient to rewrite the statistical factor, $\mathcal{F}=\expf{\xe'-\xe} n'(1+n)-n(1+n')$ (which accounts for stimulated scattering effects but neglects Bose-blocking) as
\begin{align}
\mathcal{F}
&=\mathcal{F}_1+\mathcal{F}_2
\nonumber \\
\mathcal{F}_1
&=\expf{x'-x}n'(1+n)-n(1+n')
\nonumber \\ 
\mathcal{F}_2
&=\left[\expf{\xe'-\xe}-\expf{x'-x}\right]n'(1+n).
\end{align}
The first contribution, $\mathcal{F}_1$, involves only terms related to departures of $n$ from a blackbody at the temperature $\Tz$. The second captures contributions that are to leading order only related to the differences between the electron and photon temperatures. Inserting $n=\nbb+\Delta n$ and grouping terms, we find
\begin{align}
\mathcal{F}_1&= \expf{\Delta x} f_{x'x}\,\Delta n' - f_{xx'} \,\Delta n  + (\expf{\Delta x}-1)\Delta n\Delta n'
\nonumber \\ 
&= \expf{\Delta x} (f_{x'x}+\Delta n)\,\Delta n' - (f_{xx'}+\Delta n') \,\Delta n 
\nonumber \\ 
\mathcal{F}_2&=(\expf{\Delta x\Delta \phi}-1)
\Bigg\{
f_{xx'}\frac{G(x)}{x}
+\expf{\Delta x}\left[\Delta n' + \nbb_{x'}\Delta n+\nbb_{x}\Delta n'+\Delta n\Delta n'\right]
\Bigg\}
\nonumber \\ 
f_{xx'}&=\frac{1+\nbb_{x'}}{1+\nbb_{x}}
\equiv\frac{1-\expf{-x}}{1-\expf{-x'}}
\label{eq:Fi_factors}
\end{align}
with $\Delta x=x'-x$ and $\Delta \phi=\Tz/\Te-1\equiv -\Delta \rho/\rho$ with $\rho=\Te/\Tz$. To obtain the linearized versions for $\mathcal{F}_k$, we drop all terms $\mathcal{O}(\Delta n^2)$. If in addition $\Delta \phi\ll 1$, for $\mathcal{F}_2$ one can expand $\expf{\Delta x\Delta \phi}-1\approx \Delta x\Delta \phi$. In this situation, also the terms $\mathcal{O}(\Delta \phi\Delta n)$ can be omitted.

Putting things back together, with the definitions above we obtain
\begin{align}
\left.\frac{\id  \Delta n}{\id\tau}\right|_{\rm CS}
=
\int P(x\rightarrow x',\The)\, \left\{\mathcal{F}_1(x, x')+\mathcal{F}_2(x, x')\right\}\!\id x'.
\end{align}
To convert this integro-differential equation into a set of coupled ODEs, we insert piece-wise Lagrange polynomial descriptions for $\mathcal{F}_1(x, x')$ and $\mathcal{F}_2(x, x')$ in $x'$. This ensure that exact detailed balance is maintained. The scattering kernel itself is used as a weight function to improve the convergence properties of the resultant scattering matrix. For the evolution equation of the photon occupation number around the frequency bin $x_i$, we then find\footnote{The expressions are given for centered fifth order Lagrange polynomials; however, the order can be varied inside {\tt CosmoTherm}.}
\begin{align}
\left.\frac{\id  \Delta n_i}{\id\tau}\right|_{\rm CS}
&=\sum_j S_{ij}(\Thz, \The)\left[\mathcal{F}_{1, ij}+\mathcal{F}_{2, ij}\right]
\nonumber\\
S_{ij}(\Thz,\The)&=
\int_{x_j^-}^{x_j^+} P(x_i\rightarrow x',\The)\,
\left[\sum_{k=j-2}^{j+2} L^{(k)}_j(x')\right]
\id x',
\label{eq:Dn_evolution_kernel}
\end{align}
where $\mathcal{F}_{k, ij}=\mathcal{F}_{k}(x_i, x_j)$, with $x_i$ and $x_j$ playing the roles of $x$ and $x'$ in Eq.~\eqref{eq:Fi_factors}, respectively. We furthermore have $x_j^\pm=(x_j+x_{j\pm1})/2$, defining the bin around $x_j$.

Although the frequency grid is fixed, the scattering matrix, $S_{ij}(\Thz, \The)$, has to be updated every time the electron or photon temperatures change significantly. This rather time-consuming task can be efficiently carried out using {\tt CSpack}. We attempted treating the scattering Kernel as constant within each bin, but found this to lead to problems with photon number and energy conservation at low electron temperature. This problem was remedied slightly by increasing the number of frequency grid points to resolve the width of the scattering kernel better, but in the end we found it to be beneficial to explicitly include the kernel into the weight computation.

The set of ODEs is non-linear in $\Delta n_i$ and $\Delta \rho$. To setup the Jacobian of the problem, which is required by our ODE solver, we therefore need the derivatives $\id^2 n_i/\id \Delta n_j\id\tau$ and $\id^2 n_i/\id \Delta \rho\id\tau$. The derivative with respect to $\Delta \rho$ is best evaluated numerically, using finite-differencing. We varied between centered 3- and 5-point stencils in $\Delta \rho$ but found no significant differences. The relevant terms for the derivative with respect to $\Delta n_j$ are related to 
\begin{align}
\frac{\partial\mathcal{F}_{1,ik}}{\partial \Delta n_j}
&= \left[\expf{\Delta x_{ki}} f_{ki}+\Delta n_i (\expf{\Delta x_{ki}} -1)\right]\,\delta_{jk} - \left[f_{ik}-\Delta n_k(\expf{\Delta x_{ki}} -1)\right] \,\delta_{ij}
\nonumber
\\
\nonumber
\frac{\partial\mathcal{F}_{2,ik}}{\partial \Delta n_j}&=(\expf{\Delta x_{ki}\Delta \phi}-1)
\expf{\Delta x_{ki}}\left[(1+\nbb_{x_i}+\Delta n_i)\delta_{jk} + (\nbb_{x_k}+\Delta n_k)\delta_{ij}\right],
\end{align}
where $\Delta x_{ki}=x_k-x_i$ and $f_{ik}=f_{x_i x_k}$. 
With these expressions, we can setup the jacobian of the evolution equation around a given solution. The ODE solver then iterates the solution until convergence is reached. Without non-linear terms this can be achieved in one iteration, while with non-linear terms a few iterations may be needed. The solver automatically stops the iterations once the specified relative precision (usually $\epsilon\simeq 10^{-4}-10^{-3}$) is reached.

We limit our kernel (exact) calculations to fractional energy injection to $\Delta\rho_\gamma/\rho_\gamma\lesssim 0.4$. For larger $\Delta\rho_\gamma/\rho_\gamma$, the exact solution seem unreliable. The photon number and energy densities are not conserved up to satisfactory precision and we find a electron temperature runoff. We suspect that photons start to leave high-frequency boundary at higher electron temperature, where our setup has limitations. This feature is currently under investigation and we will clarify this point in a future work. For $\Delta\rho_\gamma/\rho_\gamma\lesssim 0.4$, number and energy conservation are excellent and do not hamper the conclusions. We carried out several convergence tests and also switching off photon emission and absorption terms to confirm these statements.

\subsection{Emission and absorption terms}
Bremsstrahlung and double Compton can produce and absorb soft photons which at low frequencies drive the photon evolution towards a blackbody spectrum at the temperature of the electrons. Their contribution to the photon evolution equation is given by,
\begin{equation}
   \left. \frac{\id n}{\id\tau}\right|_{\rm e/a}=\frac{\Lambda(x)\, \expf{-\xe}}{x^3}[1-n(\expf{\xe}-1)],
    \label{eq:DC/BR}
\end{equation}
where $\xe=\phi x$, $\Lambda(x)$ is a sum of the Bremsstrahlung (BR) and double Compton (DC) terms and is a function of frequency and temperature. 
This term can be discretized trivally by replacing $n\rightarrow n_i$, implying that emission and absorption terms contribute only to the diagonal of the Jacobian matrix when $\Lambda$ is independent of $n$.

We use the standard Bremsstrahlung coefficient as in \cite{Burigana1991b,Hu1993b,Chluba2011therm} and also the more up-to date results of {\tt BRpack} \citep{BRpack}. 
For DC, we follow the work of \citet{Ravenni2020DC} and \citet{Chluba2020large}. 
In this, the non-relativistic DC emission coefficient, $\Lambda_{\rm DC}(x)$, is given by
\begin{equation}
\label{def:Lambda}
\Lambda_{\rm DC}(x)=\frac{4\rm{\alpha}}{3\pi}\Thz^2\int x^4 n(1+n) \id x.
\end{equation}
Previous works use the approximation of $n\simeq \nbb$, which inserted into the above expression implies $\int x^4 n(1+n) \id x\approx 25.976$. This is a very good approximation for small distortions but for large energy release, the full solution $n$ has to be used for accurate results.
In particular, up-scattering of photons can lead to a significant excess of photons in the Wien-tail of the CMB which means the DC emissivity can be significantly enhanced.
However, we cannot simply use Eq.~\eqref{def:Lambda} {\it as it is} in our system of equations because it will make the Jacobian matrix dense, since the emission terms would contain an integral over the full photon spectrum. To avoid this issue, we add a separate equation for the double Compton coefficient to our system of equations which preserve the sparsity of the Jacobian matrix. 

\subsection{Electron temperature equation}
\label{sec:Te-equation}
In this section, we give the expressions required to describe the electron temperature evolution. For details, we refer to \cite{Chluba2020large} but we sketch out a few details here.

The evolution of total energy density of CMB and baryons (which includes electrons) is given by,
\begin{equation}
     \frac{\id a^4\rho_{\gamma}}{a^4\id \tau}+\frac{\id a^3\rho_{\rm b}}{a^3\id \tau}+3\frac{H}{\dot{\tau}}P_{\rm b}=\frac{\dot{Q}_X}{\dot{\tau}},
     \label{eq:AAA}
 \end{equation}
where $\rho_{\gamma}$, $\rho_{\rm b}$ are CMB and baryon energy density respectively and $P_{\rm b}$ is the baryon pressure. 
The Hubble term takes into account the cooling of electrons due to the expansion of the Universe, while $\dot{Q}_X$ denotes the external heating source where we have decaying particles in mind. However, this can be replaced by any heating source in the final expression. All time derivatives are converted to optical depth derivatives using  the Thomson scattering rate, $\dot{\tau}=\Ne\,\sigma_{\rm T}\,c$.

The evolution of electron energy density and the electron temperature (which is identical to that of the baryons) is then related by,
\begin{equation}
     \frac{\id a^3\rho_{\rm b}}{a^3\id \tau}+3\frac{H}{\dot{\tau}}P_{\rm b}=\frac{3}{2} N_{\rm b}\The\left[(1+\lambda)\frac{\id{\rm ln} a \The}{\id\tau}+(1-\lambda)\frac{\id{\rm ln} a}{\id\tau}\right]
     \label{eq:BBB}
 \end{equation}
where $\lambda$ arises from relativistic temperature corrections to the total heat capacity, $C_V=\frac{3}{2}\kB N_{\rm b}(1+\lambda)$, due to the electrons:
\begin{align}
\lambda
&\approx\frac{5}{2}\,\The\frac{N_{\rm e}}{N_{\rm b}}\left[1-\frac{3}{2}\The+\frac{9}{8}\The^2\right].
 \end{align}
Here, $N_{\rm e}$ and $N_{\rm b}$ denote the electron and baryon number densities, respectively. The expression gives $\lesssim 0.5\%$ precision up to $\The\simeq 0.2$.
By combining Eq.~\eqref{eq:AAA} and \eqref{eq:BBB}, we then finally have,
\begin{equation}
     \frac{\id  \rho}{\id\tau}=\frac{\dot{Q}_X}{C_V \Thz\,\dot{\tau}}-\frac{1-\lambda}{1+\lambda}\frac{H}{\dot{\tau}}\,\rho-\frac{1}{C_V \Thz}\left.\frac{\id a^4\rho_{\gamma}}{a^4 \id\tau}\right|_{\rm CS+c/a}
     \label{eq:Tb_eq_all}
 \end{equation}
The first term in the right hand side is due to external energy injection, the second term due to Hubble cooling and the last term is due to the change in CMB energy density by to Compton scattering and emission/absorption terms, which we now turn to. The discretized version for the heat exchange due to Bremsstrahlung and double Compton from Eq.~\eqref{eq:DC/BR} and using Eq.~\eqref{eq:weights} can be written as,
\begin{equation}
    \left.\frac{\id a^4\rho_{\rm \gamma}}{a^4\id \tau}\right|_{\rm e/a}=\kappa_{z} \sum_i w_i \Lambda(x_i) \expf{-x_i\,\phi}[1-n(\expf{x_i\,\phi}-1)],
\end{equation}
 where $\kappa_{z}=\frac{8\pi h}{c^2} \left(\frac{\kB \Tz}{h}\right)^4$ (this is the factor to convert dimensionless energy density $\int x^3n \id x$ to the physical density).
 
 We next consider Compton scattering terms in both the non-relativistic case, relevant to the Kompaneets setup, and then relativistic case used in the kernel treatment.

\subsubsection{Non-relativistic regime}
The change in photon energy density for non-relativistic Compton scattering can be obtained by multiplying Eq.~\eqref{eq:Kompaneets_eq} by $x^3$ and carrying out the integral over $x$. After integrating by parts, one finds
\begin{equation}
    \left.\frac{\id a^4\rho_{\rm \gamma}}{a^4\id \tau}\right| _{\rm CS}=\kappa_{z} \left[4\The\int x^3n \id x-\Thz\int x^4n(1+n)\id x\right],
    \label{eq:CMB_density_theory}
\end{equation}
where  $n=n_{\rm pl}+\Delta n$. In the literature, one usually obtains the Compton equilibrium temperature by demanding that $\id a^4\rho_{\rm \gamma}/\id\tau=0$. This is a  reasonable assumption at high redshifts when Compton scattering happens over a short timescale compared to the Hubble rate and the photon spectrum reaches quasi-equilibrium state. However, with {\tt CosmoTherm} we fully evolve the electron temperature over time.

While Eq.~\eqref{eq:CMB_density_theory} gives the theoretical expression for the rate of change in photon energy density, in {\tt CosmoTherm} we compute it numerically using the weight factors as described in Sec. \ref{subsec:photon_distribution}. The discretized expression then becomes,
\begin{equation}
   \left. \frac{\id a^4\rho_{\rm \gamma}}{a^4\id \tau}\right| _{\rm CS}=\kappa_{z} \left[4\The\sum_i w_ix_i ^3n_i-\Thz\sum_i w_ix_i ^4n_i(n_i+1)\right],
    \label{eq:CMB_density_discrete}
\end{equation}
which then is added to Eq.~\eqref{eq:Tb_eq_all}. This does not include any Compton relativistic corrections, which can become noticeable at $\The\gtrsim 0.01$. However, most of our computations do not reach this regime.

\subsubsection{Relativistic regime}
 The equivalent expression for the kernel treatment, which includes relativistic corrections, can be obtained from Eq.~\eqref{eq:Dn_evolution_kernel} in a similar manner. The final expression is given by
\begin{align}
\left. \frac{\id a^4\rho_{\rm \gamma}}{a^4\id \tau}\right| _{\rm CS}
&=\kappa_{z}
\sum_i w_i x_i ^3\sum_j S_{ij}(\Thz, \The)\left[\mathcal{F}_{1, ij}+\mathcal{F}_{2, ij}\right]
\label{eq:CMB_density_kernel}
\end{align}
This expression allows to consistently include relativistic temperature and photon energy corrections to the energy exchange.
The terms in the Jacobian matrix can be readily obtained as discussed above. This closes the system of equations.

\subsection{Initial conditions and overall energetics}
\label{subsec:initial_condition}
To set the initial conditions with large energy release we need to solve the overall energetics of the problem including modifications to the Hubble expansion rate. The goal is to find an initial blackbody temperature such that after electromagnetic energy is injected into the background photon field, the total energy density of the standard CMB is reached. For energy injection scenarios, it is clear that the initial energy density of the photon field should be lower than that of the standard CMB, $\rho_{\rm CMB}=\rho_{\rm CMB, 0} (1+z)^4$, where $\rho_{\rm CMB, 0}\approx 0.26 \,\eV \cm^{-3}$ for $T_0=2.7255\,$~K. Similarly, we can run the calculation for large energy extraction, where the opposite is true.

For single energy release, no change to the Hubble parameter has to be included since we can start the computation with the standard Hubble parameter after the injection. In this case, the initial blackbody photon temperature is simply given by
(see =CRA20)
\begin{align}
\frac{T_{\rm in}}{\TCMB(z_{\rm in})}=
\left[
1-\DlnrhoCMB\right]^{1/4}
=
\left[
1+\Dlnrhoin\right]^{-1/4},
\end{align}
where $\DlnrhoCMBt\leq 1$ is set relative to the final CMB blackbody, while $\Dlnrhoint$ is relative to the initial CMB blackbody, implying
\begin{align}
\DlnrhoCMB
=
\frac{\Dlnrhoint}{1+\Dlnrhoint}.
\end{align}
In {\tt CosmoTherm}, we can chose the definition as required.

For the decaying scenario, small energy release does not affect the standard background evolution significantly and we can map the decaying particle lifetime to redshift ($z_X$), ignoring the change in the Hubble rate due to energy release. However, this is no longer valid for large energy release and the corresponding correction to the time-redshift relation has to be taken into account \citep{Chluba2011therm, Chluba2020large}. 

To determine the correct relations, we can use the evolution equation for the total photon energy density and the dark matter for the decaying scenario:
\begin{align}
\label{eq:rho_gamma_evol}
\frac{\id a^4 \rho_\gamma}{a^4 \id t}&=\dot{Q}_X,
\qquad
\frac{\id a^3 \rho_X}{a^3 \id t}=-\dot{Q}_X.
\end{align}
This equation assumes that the injected energy very quickly reaches the photon field and that the fractional amount of energy stored in the baryons is negligible.\footnote{It also assumes that the energy dynamics is not very sensitive to the thermalization process and exact shape of the distortion.} 
While there is no net change in energy density of combined CMB+baryon+dark matter fluid, for extremely large energy release (i.e., essentially starting with no photons in the cosmic fluid) the Hubble parameter is modified as the energy density of dark matter redshifts as $a^{-3}$ while the decay product redshift as radiation or $a^{-4}$. 
Thus, by augmenting the above equations with the first Friedmann equation we can obtain the solutions for $\rho_\gamma(a)$, $\rho_X(a)$, $H(a)$ and $t(a)$ for a given decaying particle scenario. We tabulate these solutions for later use in the main run of {\tt CosmoTherm}.

In the decaying particle scenario, we need to determine the input parameter $f_X$ [Eq.~\eqref{eq:decay_source}] to yield the correct value for the total energy release, $\Delta\rho_\gamma/\rho_\gamma$. This mapping becomes complicated as the Hubble rate is modified by energy injection due to decay while the decay rate itself depends upon the Hubble rate [Eq.~\eqref{eq:decay_source}].
We therefore solve the problem iteratively to find the correct value for $f_X$ given $\Delta\rho_\gamma/\rho_\gamma\leq 1$. 
This process converges rather quickly in just a few iterations.

\subsection{Details about the numerical implementation}
In this section we provide a few finer details about the implementation in {\tt CosmoTherm}. The basics are presented already in \citet{Chluba2011therm}, but several updates deserve mentioning.

\subsubsection{Computing the scattering matrix}
While {\tt CSpack} allows very accurate and efficient evaluation of the Compton kernel, setting up the scattering matrix is still a quite expensive step. The integrals are carried out using simple parallelization with {\tt openmp}. Since the integrals are all independent, this process scales very well with the number of cores. At lower temperatures, the scattering matrix is more sparse, such that overall fewer integrals have to be carried out, which accelerates the computations. 

However, even after parallelizing the scattering matrix setup, the computations still remains rather time-consuming. To further accelerate the computation, we implemented an interpolation scheme for the scattering matrix across temperature. Naively, only $\The$ is relevant, but since we use $x=h\nu/k\Tz$, we also need to interpolate over $\Tz$ to avoid needing to shift the frequency grid points.
For this, we predetermine a wide logarithmic grid in both $\The$ and $\Thz$ to allow for a simple 4-point polynomial interpolation. This initially defines an empty 2D array of empty matrices. Whenever we require the scattering matrix for a certain pair of $(\The, \Thz)$ on the predetermined grid (requested by the interpolation scheme) we add the corresponding matrix if it has not already been computed. In this way, we directly follow the trajectory of $(\The, \Thz)$ through the predetermined temperature grid, while leaving most pairs of temperatures never initialized. This accelerates the computations significantly, with the detailed gain depending on various settings.  
For typical settings we obtain gains in performance by more than an order of magnitude without significant loss of precision.

\subsubsection{Improving the numerical precision using intermediate shifts of the reference blackbody temperature}
For energy injection at high redshifts, the spectrum quickly decays towards a blackbody at an increased temperature. Yet, even at $z\gtrsim \pot{2}{6}$ a small $\mu$-type distortion will {\it always} be present. In order to numerically resolve this minor distortion one has to repeatedly adjust the temperature of the reference blackbody, $\Tz$. To avoid introducing terms of the type $\simeq x\partial_x n(x)\,\partial_\tau\ln a \Tz$ into the photon evolution equation, one can perform temperature shifts analytically.
Between the shifts one should use $\Tz \propto (1+z)$ and $x=h\nu/k\Tz={\rm const}$, as explained above. Thus, adjustments of the reference temperature should be carried out at various intermediate redshifts, $z_s$, using the solution at a given point and then reinitializing the ODE solver. Let us go through the individual steps.

We start with $\Tz^{(s)}=T^{(s)}_0(1+z)$ at $z_s$, where $T^{(s)}_0$ is the reference blackbody temperature at $z=0$. Considering energy injection scenarios, we usually have $T^{(s)}_0<T_{\rm CMB}=2.7255\,{\rm K}$. The total spectrum at $z_s$ is then given by
\begin{align}
n(z_s, x_s)&=\nbb(x_s)+\Delta n(z_s, x_s),
\end{align}
where $x_s=h\nu/k\Tz^{(s)}$ and $\Delta n(z_s, x_s)$ describes the distortion with respect to the reference blackbody spectrum $\nbb(x_s)=1/(\expf{x_s}-1)$, which remains constant. We now assume that at $z_s$ we have
\begin{align}
\Delta \mathcal{G}_2(z_{s})=\int x_s^2 \Delta n(z_s, x_s)\id x_s = 0,
\end{align}
which implies that the distortion does not carry any photon number contributions but only energy density contributions, i.e., $\int x_s^3 \Delta n(z_s, x_s)\id x_s\neq 0$ generally. There is no assumption about the amplitude of the distortion.

We then evolve the spectrum to a new redshift $z_{s+1}$. This is done using {\tt CosmoTherm} and includes all evolution effects. The redshift step, $\Delta z=z_{s+1}-z_{s}$ need not be small, but just assumes that $T^{(s)}_0$ is kept constant during the evolution. Due to DC and BR emission, part of the distortion will thermalize such that at $z_{s+1}$ we have
\begin{align}
\Delta \mathcal{G}_2(z_{s+1})=\int x_s^2 \Delta n(z_{s+1}, x_s)\id x_s \neq 0.
\end{align}
We can now decide to redefine the reference blackbody temperature to $T^{(s+1)}_0$, such that 
\bsub
\begin{align}
n(z_{s+1}, x_{s+1})&=\nbb(x_{s+1})+\Delta \tilde{n}(z_{s+1}, x_{s+1})\equiv n(z_{s+1}, x_s)
\\
\label{eq:cond_number}
\Delta \mathcal{\tilde{G}}_2(z_{s+1})&=\int x_{s+1}^2 \Delta \tilde{n}(z_{s+1}, x_{s+1})\id x_{s+1}=0.
\end{align}
\esub
Explicitly, we have
\begin{align}
\label{eq:Dntilde}
\Delta \tilde{n}(z_{s+1}, x_{s+1})&=
\nbb(x_{s})-\nbb(x_{s+1})+\Delta n(z_{s+1}, x_s)
\end{align}
and hence the condition
\begin{align}
\int x_{s+1}^2 \left[\nbb(x_{s+1})-\nbb(x_{s})\right]\id x_{s+1}
&=
\int x_{s+1}^2\Delta n(z_{s+1}, x_s)\id x_{s+1}
\end{align}
to ensure Eq.~\eqref{eq:cond_number}. This then implies
\begin{align}
f^3=\left(\frac{T_0^{(s+1)}}{T_0^{(s)}}\right)^3&=1+\frac{\Delta \mathcal{G}_2}{\mathcal{G}_2^{\rm pl}}.
\end{align}
To restart the solver with the new initial distortion, Eq.~\eqref{eq:Dntilde}, we have two possibilities: i) we reuse the same grid of frequency points, $x_s$, even after the shifting of the reference temperature or ii) we introduce a new grid of points with shifted frequency bins. With method i) we need to interpolate the solution for $\Delta n(z_{s+1}, x_s)$ at $x^*=x_s f$. This causes a problem at one of the boundaries of the frequency domain, requiring one to extrapolate the solution. If the temperature shifting procedure is repeated frequently, this is only a minor issue. On the other hand, for method ii), introducing a new grid of points can be expensive, as all the discretization variables (e.g., the integral weights) have to be recomputed. However, in this case, the solution remains numerically highly accurate.

As in the original version of {\tt CosmoTherm}, we will follow i) in our computations. However, we no longer assume that the temperature shift remains small. To improve the numerical stability we write
\begin{align}
\nbb(x_{s})-\nbb(x_{s+1})&
=\frac{\expf{-x_s}[1-\-\expf{-(x_{s+1}-x_{s})}]}{(1-\expf{-x_s})(1-\expf{-x_{s+1}})}
\end{align}
and define the function $g(x)=1-\expf{-x}$, treating it by using a Taylor series at $x\ll 1$. We then finally have
\begin{align}
\Delta \tilde{n}(z_{s+1}, x_{s})&=
\frac{g(x_{s}-x^*)}{g(x^*)g(x_s)}\,\expf{-x^*}+\Delta n(z_{s+1}, x^*)
\end{align}
with $x^*=x_s f$. For $f>1$, the solution has to be extrapolated at the upper boundary, while for $f<1$ it is the lower boundary.

\subsubsection{{\label{subsec:temp_shifting}}Criterion for temperature shifting}
While we have explained how to shift the reference temperature and restart the computation of the distortion evolution, we still need to decide how to best determine when the shift is needed. This can be easily achieved by comparing the total energy density carried by $\Delta n$ with that carried by the pure distortion part, which we define using the criterion $\int x^2 \Delta n_{\rm d} \id x=0$. Using the solution $\Delta n(z, x)$, the total  fractional energy density carried by these photons relative to the blackbody part at temperature $\Tz$ is given by
\begin{align}
\frac{\Delta \rho_\gamma}{\rho_\gamma^{\rm pl}}\Bigg|_{\rm tot}
&=\frac{\Delta \mathcal{G}_3}{\mathcal{G}_3^{\rm pl}}.
\end{align}
Given the effective temperature of the photon field based on the total number of photons
\begin{align}
T_N
&=\Tz\left(1+\frac{\Delta \mathcal{G}_2}{\mathcal{G}_2^{\rm pl}}\right)^{1/3}
\end{align}
we can then isolate the energy density contribution from the distortion part, $\Delta n_{\rm d}$, of the spectrum
\begin{align}
\frac{\Delta \rho_\gamma}{\rho_\gamma^{\rm pl}}\Bigg|_{\rm d}
&=1+\frac{\Delta \mathcal{G}_3}{\mathcal{G}_3^{\rm pl}}-\left(1+\frac{\Delta \mathcal{G}_2}{\mathcal{G}_2^{\rm pl}}\right)^{4/3}.
\end{align}
Conversely, the energy density contribution from the part which carries photon number (i.e., $\int x^2 \Delta n_N \id x\neq 0$) is given by
\begin{align}
\frac{\Delta \rho_\gamma}{\rho_\gamma^{\rm pl}}\Bigg|_N
&=\left(1+\frac{\Delta \mathcal{G}_2}{\mathcal{G}_2^{\rm pl}}\right)^{4/3}-1.
\end{align}
To decide when to reset the reference temperature and restart the calculation can therefore use the simple criterion
\begin{align}
\frac{\Delta \rho_\gamma}{\rho_\gamma^{\rm pl}}\Bigg|_N
\leq \epsilon\times\frac{\Delta \rho_\gamma}{\rho_\gamma^{\rm pl}}\Bigg|_{\rm tot}
\end{align}
with $\epsilon \simeq 0.01-0.1$. This ensures that the distortion part of the solution is always numerically highly resolved and dominates $\Delta n$.
In particular for large energy release cases, this improved setup was required to obtain accurate results for the distortion visibilities.

\section{CMB spectral distortion solutions for large energy release and extraction}
\label{sec:spec_dist_soln}

\begin{figure}
\centering 
\includegraphics[width=0.98\columnwidth]{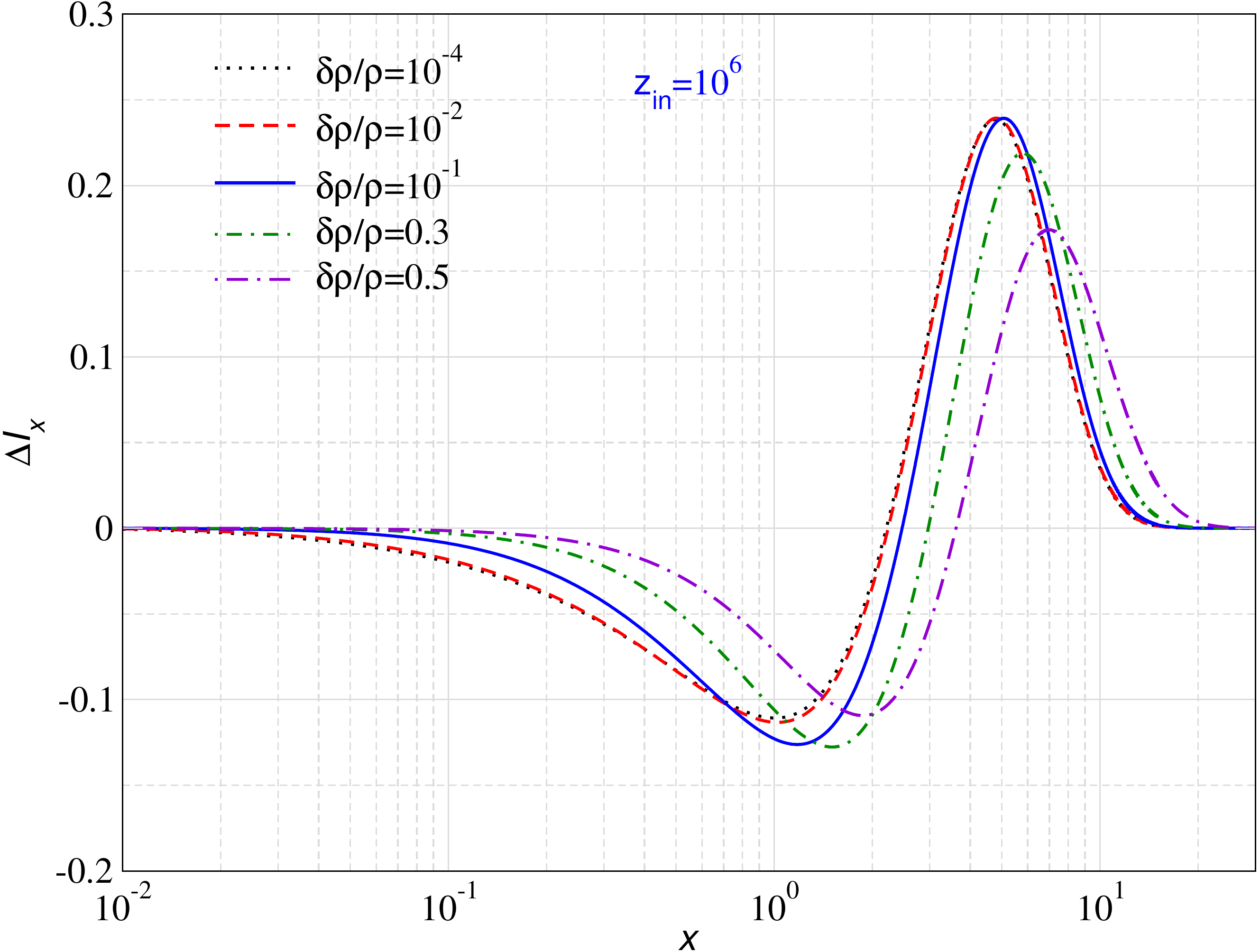}
\\
\caption{Intensity ($\Delta I_x=x^3\Delta n$) as a function of $x$ for different fractional energy injections, $\DlnrhoCMBt$, from non-linear calculations for $z_{\rm in}=10^6$ using the Kompaneets treatment. To highlight the differences in the distortion shape, the curves for each case are re-scaled such that $\int x^3\Delta n dx=1$.}
\label{fig:distortion_z_1e_6}
\end{figure}

\begin{figure}
\centering 
\includegraphics[width=0.98\columnwidth]{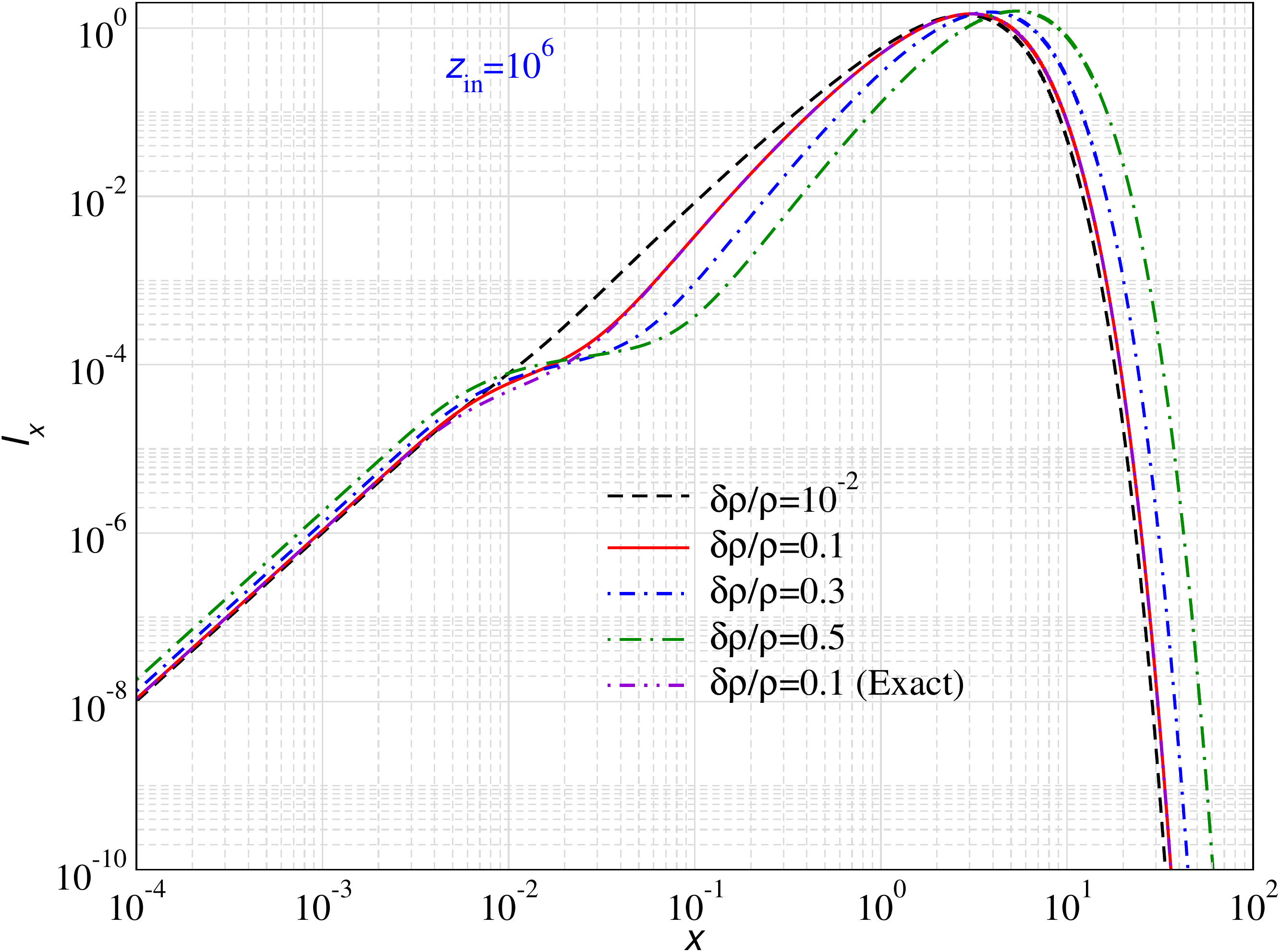}
\\
\caption{Intensity of the total photon spectrum at $z=0$ using the non-linear Kompaneets setup. The injection redshift is $z_{\rm in}=10^6$. For $\DlnrhoCMBt=0.1$, we also show the result of the scattering kernel treatment.}
\label{fig:total_n}
\end{figure}
In this section, we carry out a comparative study of spectral distortion shapes and show the limitation of the linearized PDE treatment. For simplicity, we focus on single energy release ($\delta$-function in redshift), but the conclusions remain similar for decaying particles.

In Fig. \ref{fig:distortion_z_1e_6}, we present the CMB spectral distortion shape as a function of fractional energy injections at injection redshift $z_{\rm in}=10^6$. For $\DlnrhoCMBt\ll 1$, we obtain the standard $\mu$-distortion solution, as expected. However, for large energy release ($\DlnrhoCMBt \gtrsim 0.01$), the distortion shapes become a function of $\DlnrhoCMBt$, as expected due to non-linear nature of equations. In particular, the distortion shows an enhancement of the Wien-tail photon number and a general shift of the spectrum towards higher frequencies.
The main trends are even more evident in Fig.~\ref{fig:total_n}, where we show the total CMB spectrum for comparable cases. 
For larger $\DlnrhoCMBt$, the electron temperature is higher, which boost the CMB photons to higher energy. At low frequencies, photon non-conserving processes establish a Planckian spectrum which lies at a higher temperature due to energy injection to CMB. Visually, it can be seen that there is a higher probability for the distortions to survive until today at higher $\DlnrhoCMBt$. This is because the thermalization process becomes less efficient as the distortion becomes larger \citep{Chluba2020large}.  

In Fig.~\ref{fig:total_n}, we also show the result obtained with the full kernel treatment and $\DlnrhoCMBt=0.1$. The solution agrees very well with that obtained from the non-linear Kompaneets treatment. Overall, corrections at the level of a few percent are expected \citep{Chluba2020large}, which we confirm here.

\begin{figure}
\centering 
\includegraphics[width=0.98\columnwidth]{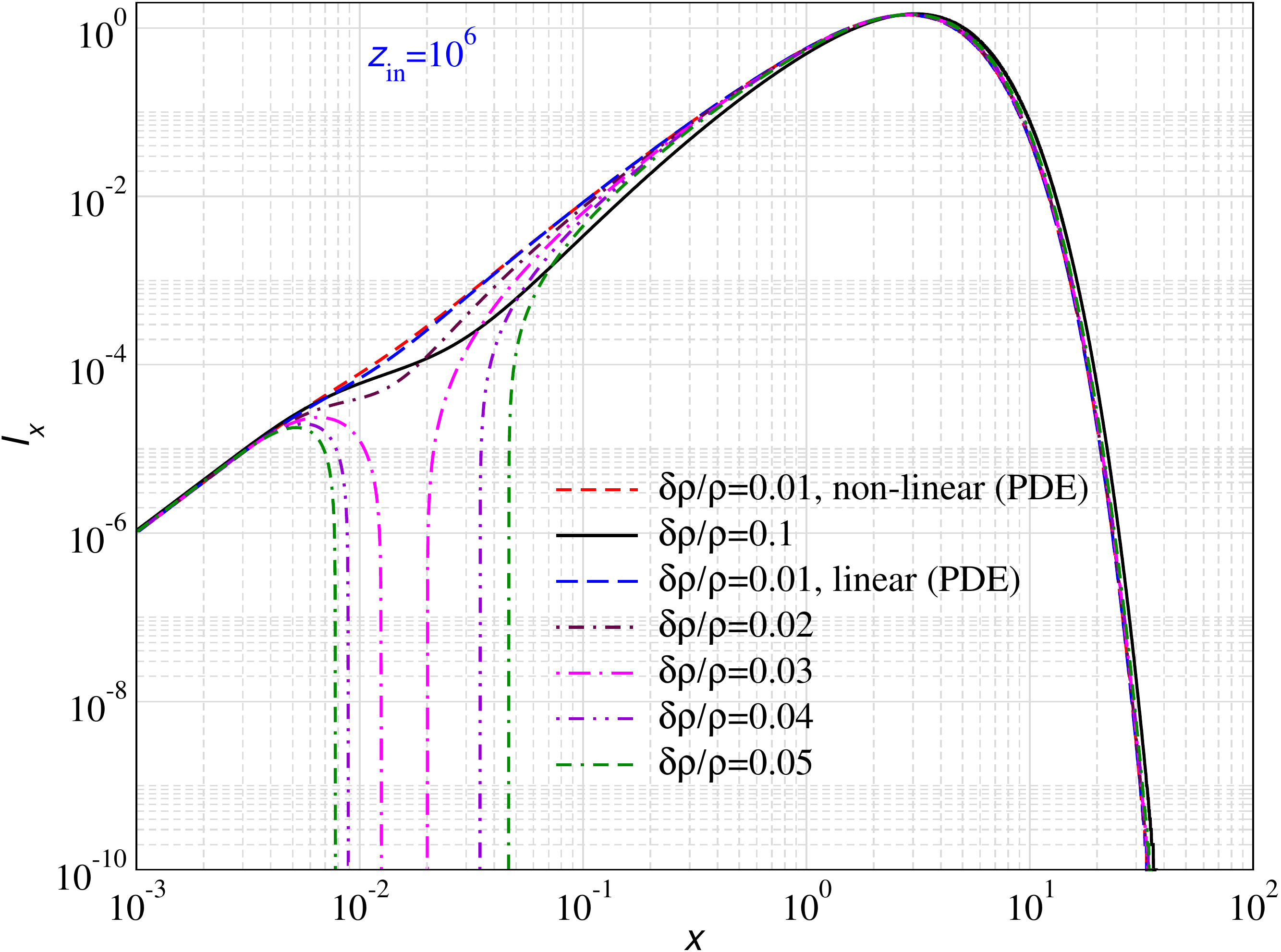}
\\
\caption{Final distorted CMB spectrum obtained from {\tt CosmoTherm} for energy injection cases as shown in the plot at injection redshift $z_{\rm in}=10^6$. The small distortion approximation produces unphysical features at low frequency which are absent in the non-linear solution.}
\label{fig:spectrum_linear_comp}
\end{figure}

\vspace{-3mm}
\subsection{Comparison of linear and non-linear solutions}
\label{sec:soln_comp}
In this section, we highlight the shortcomings of the linearized Kompaneets treated. In Fig.~\ref{fig:spectrum_linear_comp}, we compare these solutions for few single energy injection cases using the non-linear and linearized Kompaneets treatment. We can already see differences between the two solutions at $\DlnrhoCMBt=0.01$. 
The linear solution develops unphysical (negative) features in $I_x=x^3 n$ at larger values of $\DlnrhoCMBt$. It can be explained as follows: The total spectrum is given by, $n=n_{\rm{pl}}+\Delta n$. In the linear treatment, the distortion $\Delta n$ is negative for $x\lesssim 2.7$, as most of the CMB photons in this range are upscattered. The spectrum $n$ starts to become negative once $\Delta n$ starts to dominate the background CMB Planckian spectrum. Initially, this negative feature is narrow. However, as we increase  $\DlnrhoCMBt$, $\Delta n$ dominates over $n_{\rm{pl}}$ even further and the negative feature widens in frequency. Indeed we see this aspect as shown in Fig.~\ref{fig:spectrum_linear_comp}.  

We can obtain an estimate of the critical value of $\DlnrhoCMBt$ at which the linear treatment becomes unphysical. This happens when the CMB intensity is equal to the intensity of distortion spectrum.  We can do a simple estimate assuming the distortion to be that of a standard $\mu$-distortion which is produced shortly after energy injection. 
For the spectrum of a $\mu$-distortion we have
\begin{equation}
M(x)\approx  \expf{-\frac{x_{\rm c}}{x}} \,\frac{\expf{x}}{(\expf{x}-1)^2}\left[\frac{x}{\beta}-1\right],
\end{equation}
where $x_{\rm c}\approx\pot{8.6}{-3}\sqrt{(1+z)/\pot{2}{6}}$ is the critical frequency, which is determined by the competition of Compton and double Compton \citep{Hu1993,Chluba2015GreensII} and $\beta\approx 2.19$. The factor $\expf{-\frac{x_{\rm c}}{x}}$ takes into account that at low frequencies the spectrum returns to a blackbody. 
Since the frequency at which the solution will first become unphysical is at $x\ll 1$, we can take the low-frequency limit of the equations, which means $M(x)\approx -\expf{-\frac{x_{\rm c}}{x}}/x^2$. For the Planckian, we similarly have $\nbb\approx 1/x$.
We then obtain the condition
\begin{equation}
    \frac{1}{x}\approx 1.4\times\DlnrhoCMB\times \frac{\expf{-x_{\rm c}/x}}{x^2},
\end{equation}
where we used the simple estimate $\mu\approx 1.4\,\DlnrhoCMBt$, which should be valid right after the injection. 
To determine the critical value for $\DlnrhoCMBt$ at a given redshift, we need to also find the frequency at which the condition is first fulfilled.
Replacing $x\rightarrow \xi x_{\rm c}$, we have the modified condition $\xi\expf{1/\xi}\equiv 1.4\DlnrhoCMBt/x_{\rm c}$. Since the function $\xi\expf{1/\xi}$ has a minimum at $\xi=1$, this can only be fulfilled if 
\begin{equation}
\DlnrhoCMBt \lesssim
\frac{{\rm e}\times x_{\rm c}}{1.4} 
\simeq 1.9\,x_{\rm c}.
\label{eq:rho_crit}
\end{equation}
For $z_{\rm in}=10^6$, this means $\DlnrhoCMBt\approx 0.012$. The value for the critical energy release scales only weakly with $z_{\rm in}$, since $x_{\rm c}\propto \sqrt{z_{\rm in}}$. At $z\lesssim \pot{2}{5}$, one would furthermore using the BR critical frequency $x_{\rm c}\simeq \pot{1.23}{-3}[(1+z)/\pot{2}{6}]^{-0.672}$ \citep{Chluba2015}.

We confirmed the above estimate numerically using {\tt CosmoTherm}. Indeed for $\DlnrhoCMBt=0.011$ and $z_{\rm in}=10^6$ we found that the solution remained 'physical' during the full evolution, while for $\DlnrhoCMBt\gtrsim 0.012$ it developed a 'negative' photon occupation around $x\simeq x_{\rm c}$. We also mention that even if the solution becomes unphysical during part of the evolution, it can return to a state that appears physical once thermalization processes reduce the amplitude of the distortion. Hence, just inspecting the solution at the final stage does not ensure that at some intermediate stage non-linear terms remained small. Equation~\eqref{eq:rho_crit}, provides a fairly robust condition that should not be violated at {\it any} stage of the evolution.

\subsection{Large energy extraction}
\label{sec:single_energy_extraction}

\begin{figure}
\centering 
\includegraphics[width=\columnwidth]{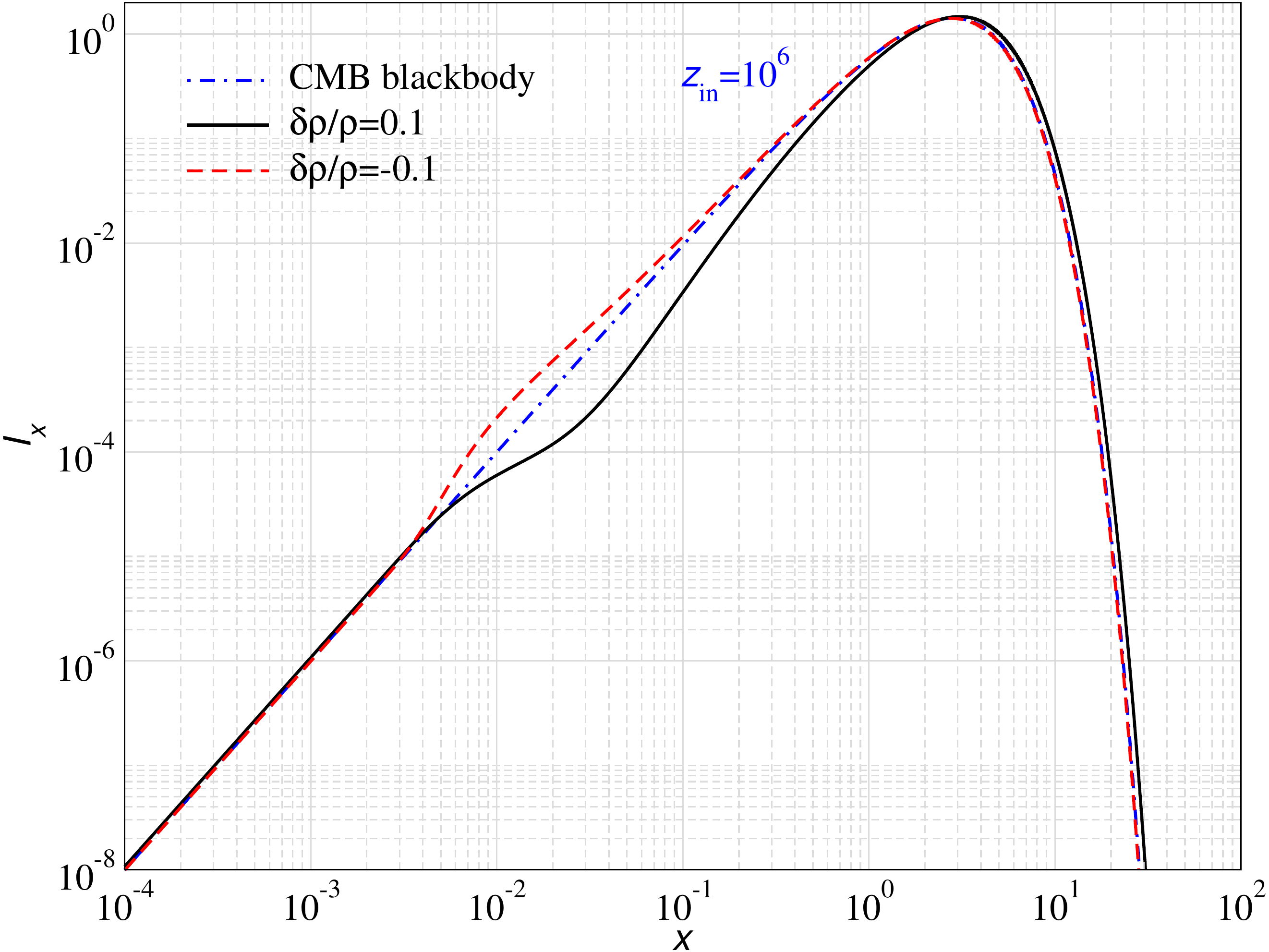}
\\
\caption{Distorted CMB spectrum for energy injection and comparison with energy extraction for $\DlnrhoCMBt=\pm 0.1$ at $z_{\rm in}=10^6$.}
\label{fig:negative_Drho}
\end{figure}

\begin{figure}
\centering 
\includegraphics[width=\columnwidth]{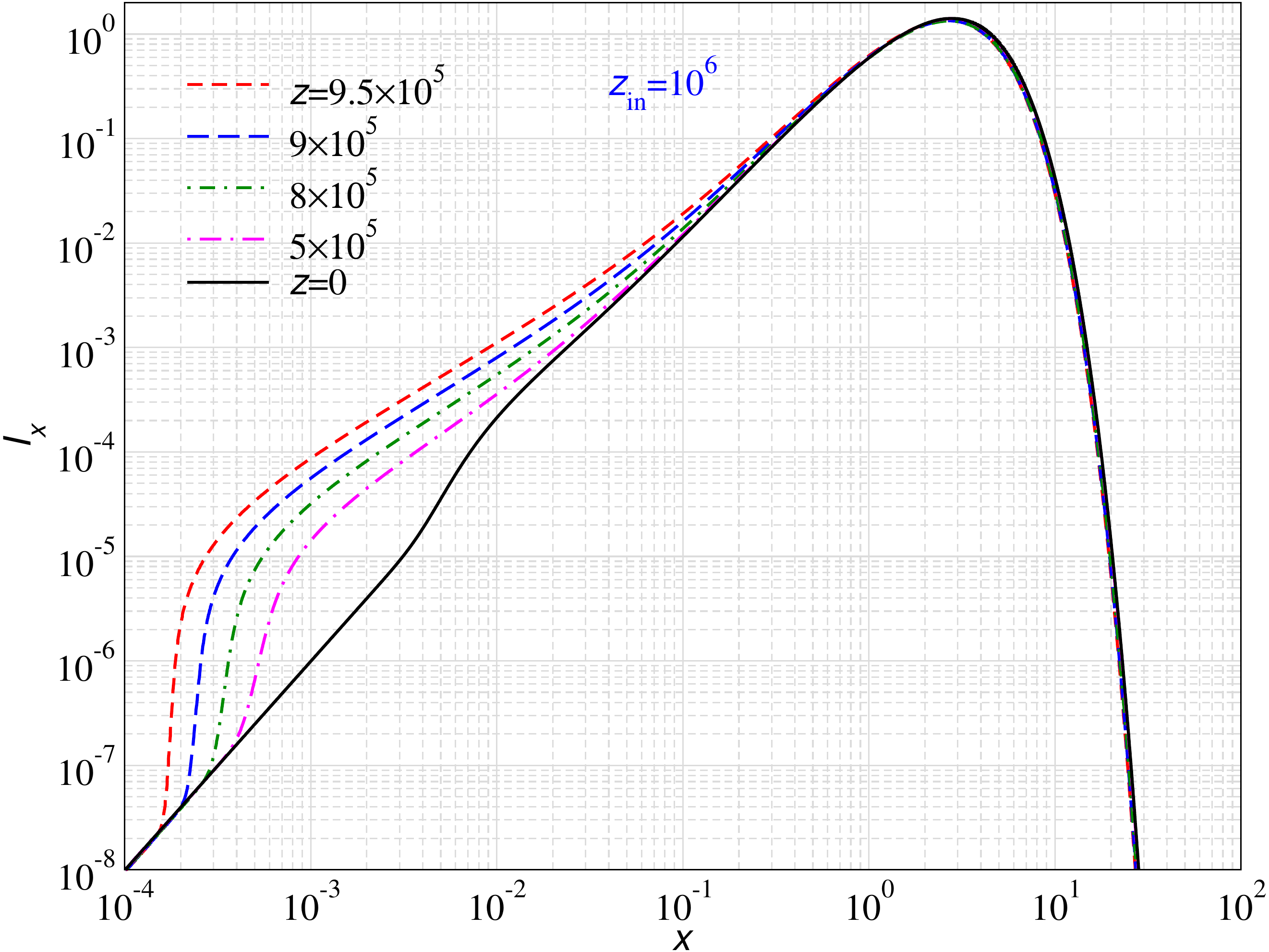}
\\
\caption{Snapshot of distorted CMB spectrum for $\DlnrhoCMBt=- 0.1$ for a few different redshifts.}
\label{fig:snapshot_Drho}
\end{figure}

While we mostly deal with energy injection to the CMB in this work, we also illustrate the CMB spectral distortion solutions from {\it energy extraction} using this framework. Exotic processes such as dark matter-baryon interaction can in principle cool the background baryons/electrons which causes an energy extraction process from the CMB \citep{Yacine2015DM, Yacine2021}. 
The same effect occurs with baryons \citep{Chluba2005, Chluba2011therm, Khatri2011BE}, but there the distortion remains very small and linear.
We model such situation as negative energy injection. An example is shown in Fig.~\ref{fig:negative_Drho}. Electron cooling results in negative $\mu$-type distortion as the CMB photons expend their energy to keep the electrons at the same temperature as that of CMB.
This results in accumulation of photons at low frequency as opposed to missing photons in the case of energy injection. 
This affects the region in which DC and BR emission is relevant and therefore should change the distortions visibility. Since the emission region moves towards lower frequencies, and because the thermalization optical depth scales as $\tau_\mu\simeq \int \Thz x_{\rm c} \id \tau$ \citep[e.g.,][]{Chluba2015}, one expects the distortions to thermalize more slowly than for the corresponding energy injection case. However, a more detailed discussion is left to future work.

We would also like to highlight how the extraction of energy leads to a shock-like structure in the low-frequency spectrum. This is illustrated in Fig.~\ref{fig:snapshot_Drho}, where we also give snapshots of the distortion solution at a few intermediate redshifts. The distortion shows a steep pile-up of photons at $x\simeq 10^{-3}$ caused by cooling and condensation of photons \citep{Levich1969}. Numerically, it is therefore difficult to treat these cases over a wide range of parameters. For the corresponding energy injection case, the distortion shape is hardly modified between the two snapshots, which highlights the difference in the physics of these two situations.

\vspace{-3mm}
\section{Spectral distortion constraints on single energy injection scenarios}
\label{sec:single_energy_case}
In this section, we derive constraints on single energy injections from CMB spectral distortion using \COBEF \citep{Fixsen1996} data. We obtain these by computing the visibility function for the distortions, which is defined as the probability of the distortions to survive until today, given the energy was injected at redshift $z_{\rm h}$. The distortion visibility function can then be written as,
\begin{equation}
\zeta(z)=\frac{\frac{\Delta \rho_\gamma}{\rho_\gamma}\big|_{\rm d}}{\frac{\Delta \rho_\gamma}{\rho_\gamma}\big|_{\rm tot}}.
\end{equation}
The denominator is determined by the total energetics of the problem, while for the numerator we need the final solution with appropriately defined CMB reference temperature (see Sec. \ref{subsec:temp_shifting} for additional discussion).

\begin{figure}
\centering 
\includegraphics[width=\columnwidth]{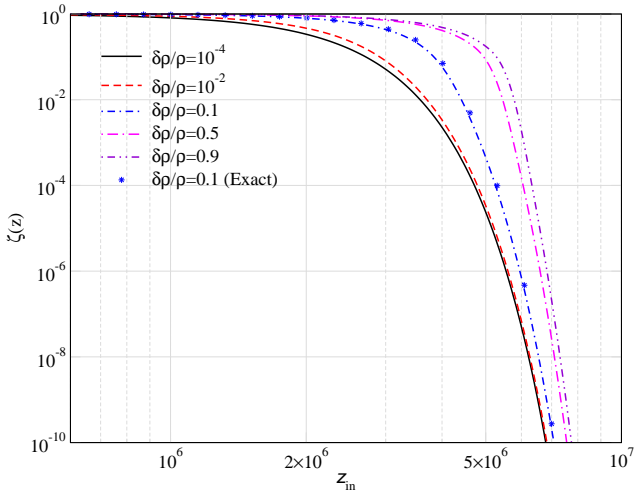}
\\
\caption{Visibility function as a function of redshift for different energy injection as shown in the figure. The solutions are obtained from the non-linear Kompaneets treatment. For $\DlnrhoCMBt=0.1$, we also show the result from the kernel treatment finding good agreement between the two.}
\label{fig:visibility_delta}
\end{figure}

\begin{figure}
\centering 
\includegraphics[width=\columnwidth]{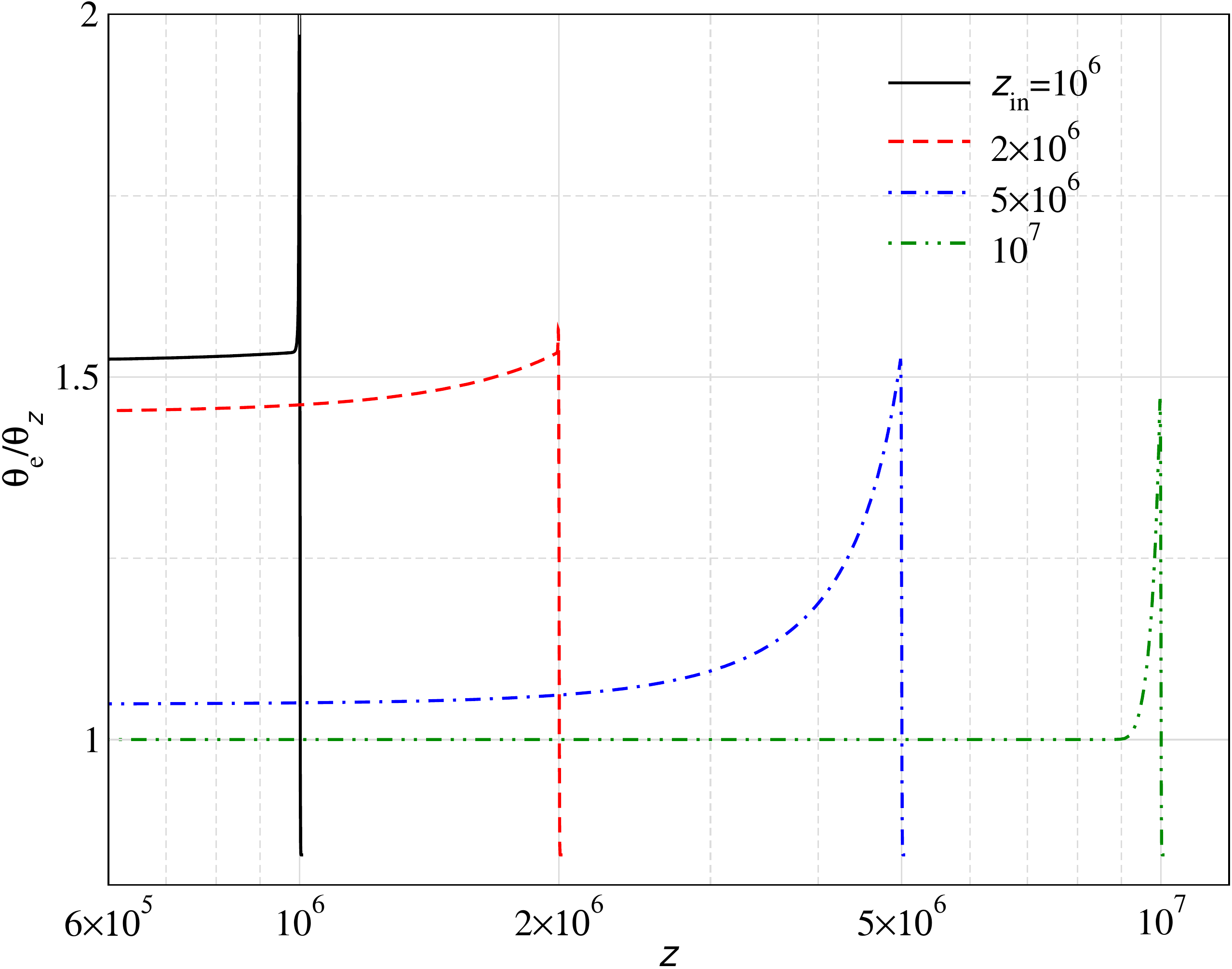}
\\
\caption{Evolution of electron temperature as a function of $z$ for single energy injection cases with $\DlnrhoCMBt(z_{\rm in})=0.5$.
For comparison, we also show the target CMB temperature. For late large energy injection $\Te$ remains larger that $\TCMB$, as thermalization becomes incomplete.}
\label{fig:T_e_delta}
\end{figure}

\begin{figure}
\centering 
\includegraphics[width=\columnwidth]{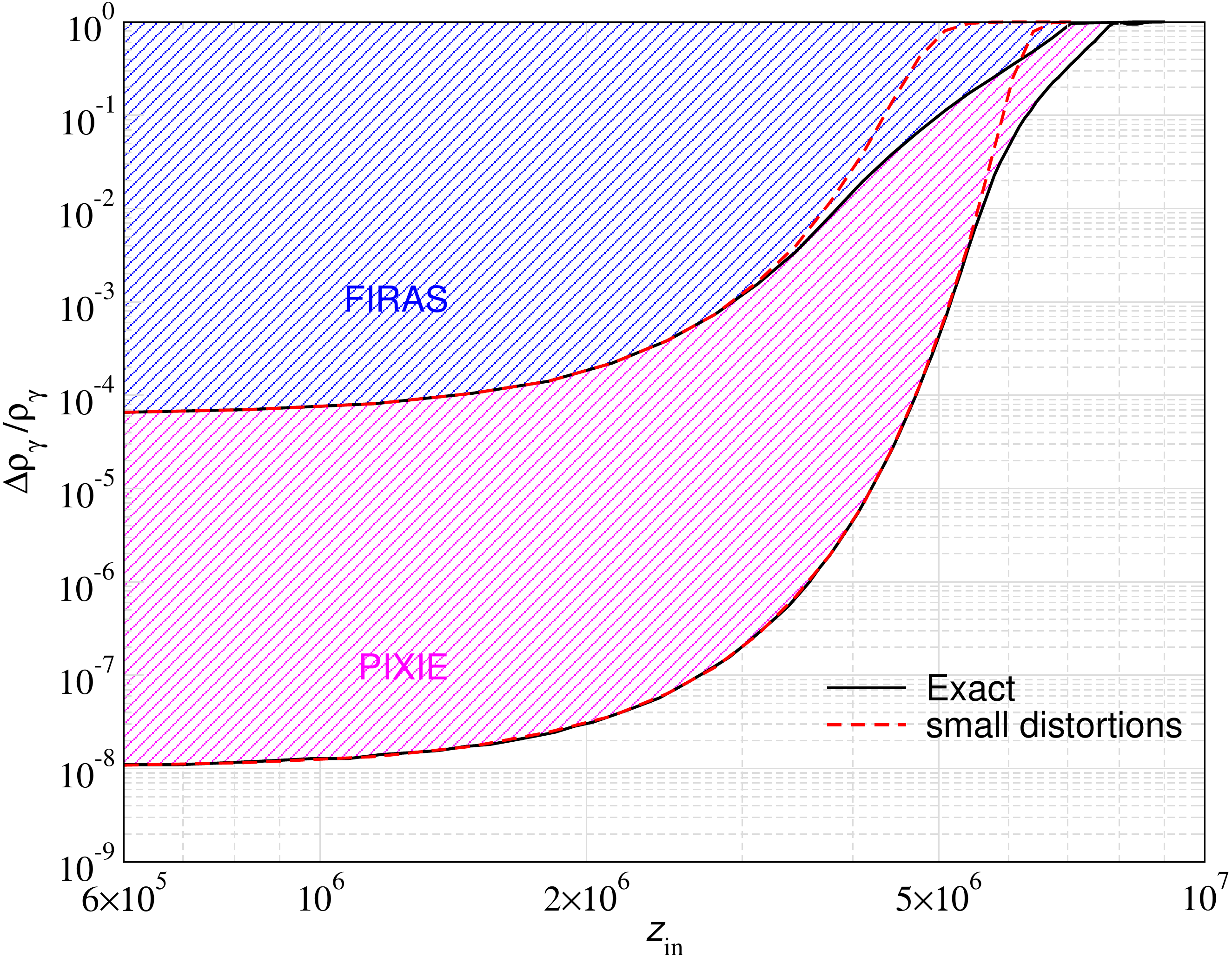}
\\
\caption{Constraints on single energy injection scenarios.
The blue region is excluded by \COBEF, while the purple region could be excluded with a future spectrometer similar to \PIXIE.}
\label{fig:constraints_delta}
\end{figure}

In Fig. \ref{fig:visibility_delta}, we illustrate the visibility function for a few representative cases. All cases are computed using fixed values of $\DlnrhoCMBt$ as annotated.
For $z_{\rm in}\lesssim 10^6$, the distortion visibility is close to unity as photon non-conserving processes (i.e., double Compton and Bremsstrahlung) become inefficient \citep{Danese1982, Burigana1991, Hu1993, Chluba2011therm, Khatri2012b}. At higher redshifts, the photon non-conserving processes in combination with Compton scattering tend to restore distorted CMB spectrum to a Planckian spectrum. This manifests in an exponentially decreasing $\zeta(z)$ in Fig.~\ref{fig:visibility_delta}. There is a clear tendency for the distortion to survive longer at large energy release. This is due to a large increase of the electron temperature, which changes the relative importance of Compton scattering and recoil compared to small energy release. 
The DC emissivity is also on average lower because until the final stages of the evolution, $\Thz^2 \int x^4 n(1+n)\id x<\theta_{\rm CMB}^2\,\int x^4 \nbb(1+\nbb)\id x$. A combination of these effects leads to a decrease in the effective photon emissivity, making the thermalization less efficient (see CRA20). 

Our results are in agreement with CRA20. However, we want to remind the reader that in this paper, we are evolving the distorted photon spectrum in a fully time-dependent manner, while CRA20 used a quasi-stationary approximation, ignoring the time-dependence. Since at these high redshifts, collision processes are extremely fast compared to the Hubble rate, the assumption that the distorted CMB spectrum evolves along a series of quasi-stationary solutions is reasonable, which we further verify here. 
In our non-linear Kompaneets treatment, we also do not account for relativistic corrections to the Compton process. These do change the results for the visibility by a small amount, as one can appreciate from the case for $\DlnrhoCMBt=0.1$ obtained with the full kernel treatment in Fig.~\ref{fig:visibility_delta}. However, the overall corrections are limited to $\simeq 10\%-20\%$ as also argued in CRA20, which we shall neglect below.  

For illustration, we also show the evolution of electron temperature in case of large energy release in Fig.~\ref{fig:T_e_delta}. %
We can observe a huge boost in electron temperature at the instant of energy injection. This boost persists for $z_{\rm in}\lesssim 10^6$ due to the inefficiency of photon non-conserving processes at these redshifts, with a large distortion being frozen in at high frequencies. At $z_{\rm in}\gtrsim 10^7$, there is a boost followed by a steep fall as the electrons cool by emitting soft photons which leads to thermalization of the distortions. In this case, $\Te\simeq \TCMB$ at the end of the evolution. For the shown scenarios, relativistic correction are expected to become noticeable for $z_{\rm in}\gtrsim \pot{5}{6}$ (CRA20).

We can now constrain the allowed value of $\DlnrhoCMBt$ by requiring that the surviving CMB distortion today to be $\lesssim 10^{-5}$. In this work, we use the criteria that $\DlnrhoCMBt\lesssim 6\times 10^{-5}$ \citep{Fixsen1996}. 
We also perform the estimate for a future CMB spectrometer similar to \PIXIE that could reach $\DlnrhoCMBt\lesssim 10^{-8}$ \citep{Kogut2016SPIE, Chluba2021ExA}.
The corresponding constraints on the energy injection are shown in Fig.~\ref{fig:constraints_delta} for the non-linear Kompaneets treatment.
For reference, we compare the constraints from large energy release and compare with the small distortion approximation, which is typically used in literature. Clear differences can be seen at higher redshifts. As the visibility reduces at high redshifts allowing more energy to be injected to the existing CMB, the non-linear effects start to manifest and there is a significant departure from the small distortions limit.  
Our results are in very good agreement with CRA20, however, here we converted the distortion into a limit on $\DlnrhoCMBt\leq 1$ rather than $\Dlnrhoint$, which was used there.

\section{Energy injection from decaying particles}
\label{sec:decay}

\begin{figure}
\centering 
\includegraphics[width=\columnwidth]{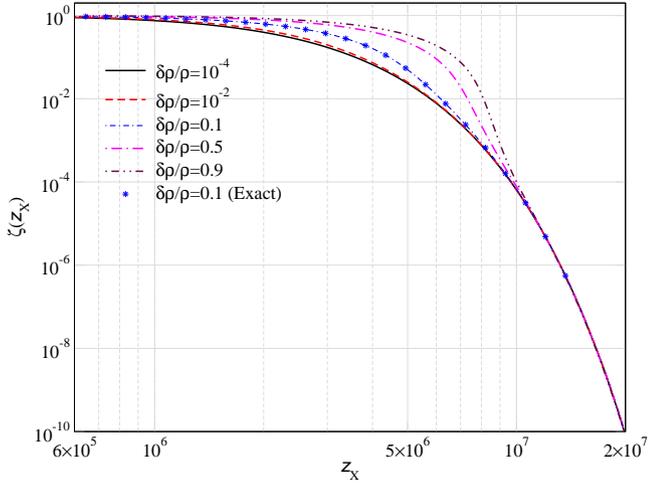}
\\
\caption{Visibility function as a function of lifetime for decay with different total energy injection as shown in the figure including the modification to Hubble. We also show a representative case with the exact calculation. }
\label{fig:visibility_decay}
\end{figure}


\begin{figure}
\centering 
\includegraphics[width=\columnwidth]{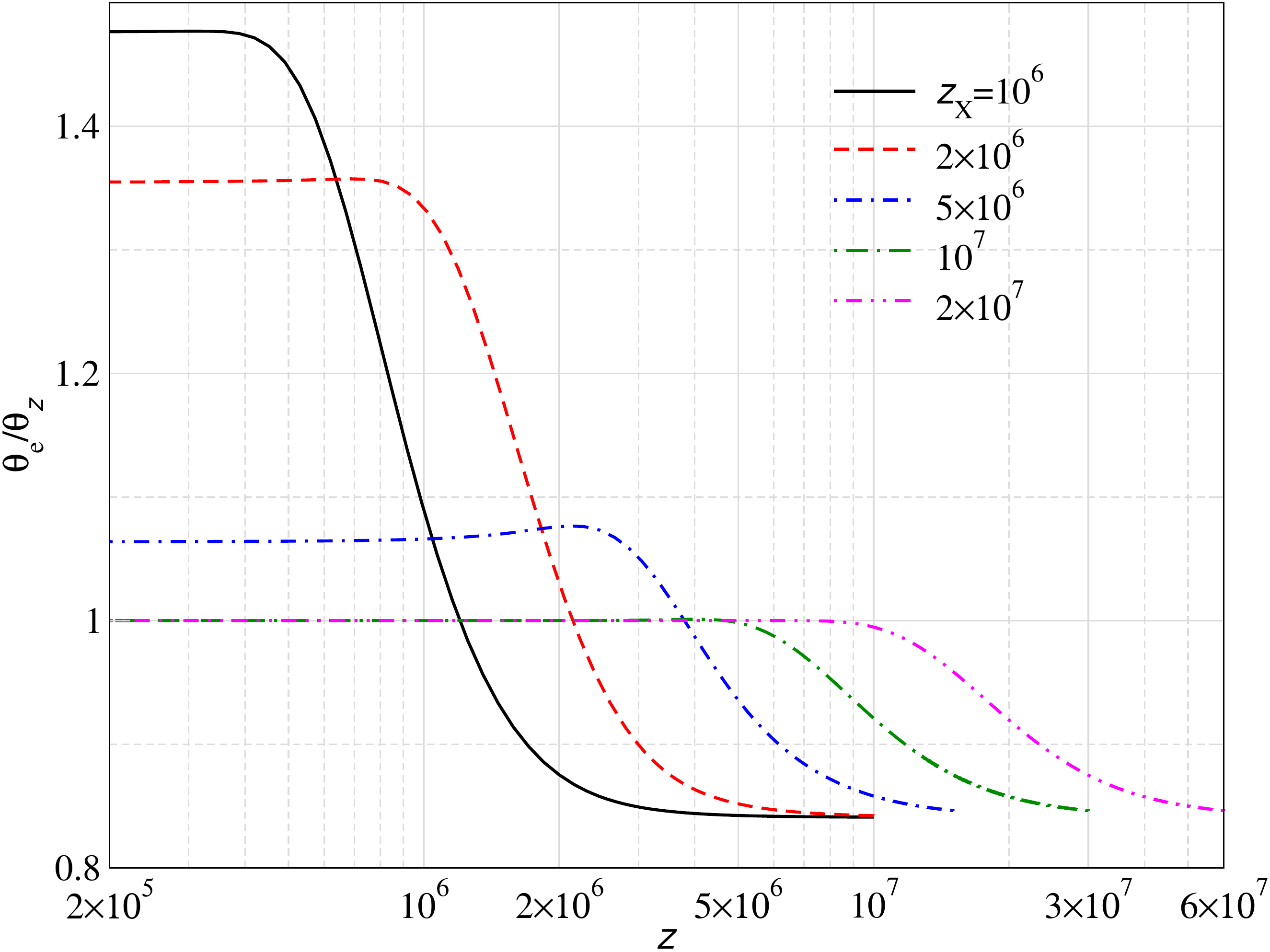}
\\
\caption{Evolution of electron temperature as a function of $z$ for decaying particles with $\DlnrhoCMBt=0.5$ and  lifetime redshift as shown in the plot.}
\label{fig:T_e_decay}
\end{figure}


\begin{figure}
\centering 
\includegraphics[width=\columnwidth]{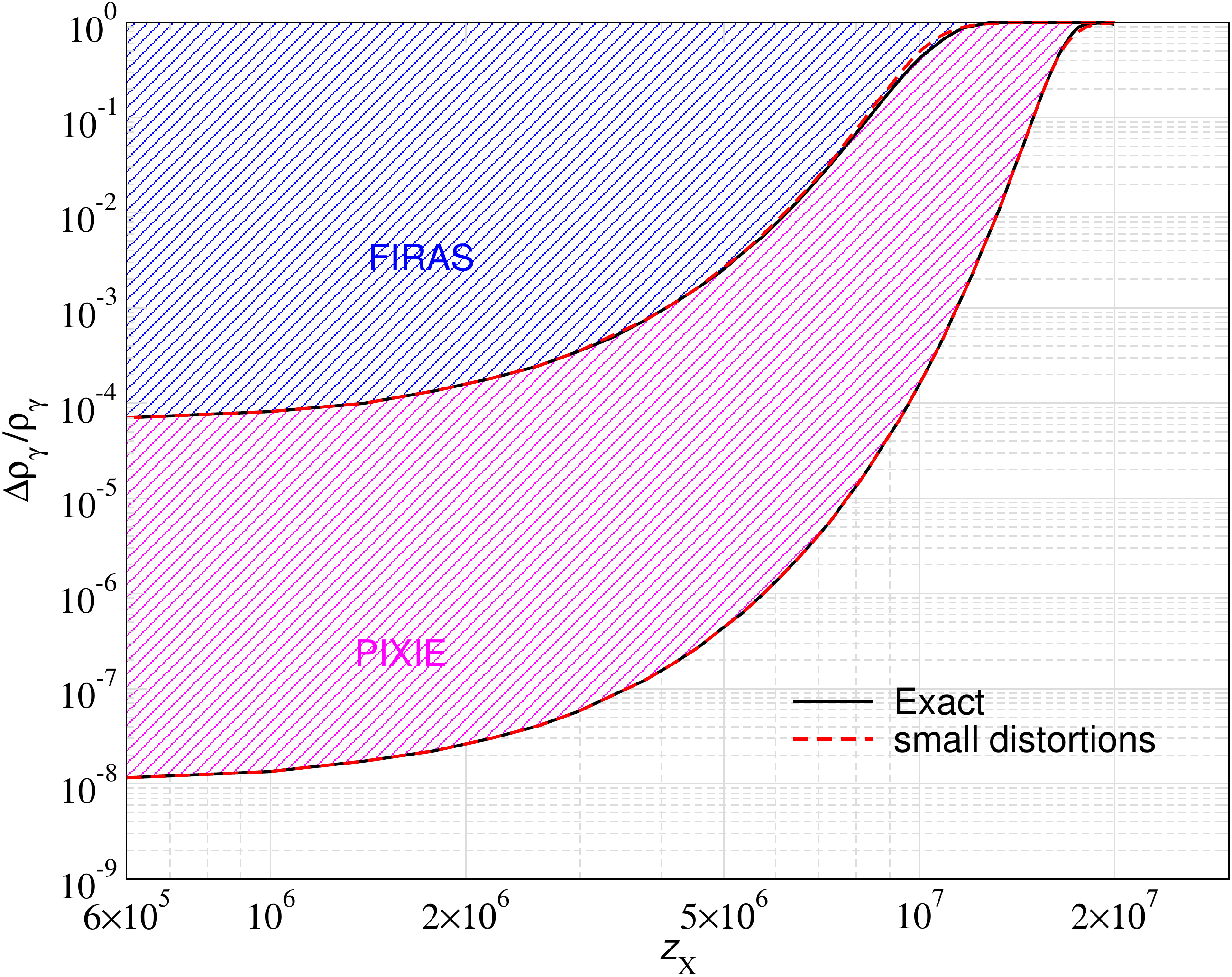}
\\
\caption{Constraints on decaying particle  energy injection scenarios.
The blue region is excluded by \COBEF, while the purple region could be excluded with a future spectrometer similar to \PIXIE.}
\label{fig:constraints_decay}
\end{figure}


\begin{figure}
\centering 
\includegraphics[width=\columnwidth]{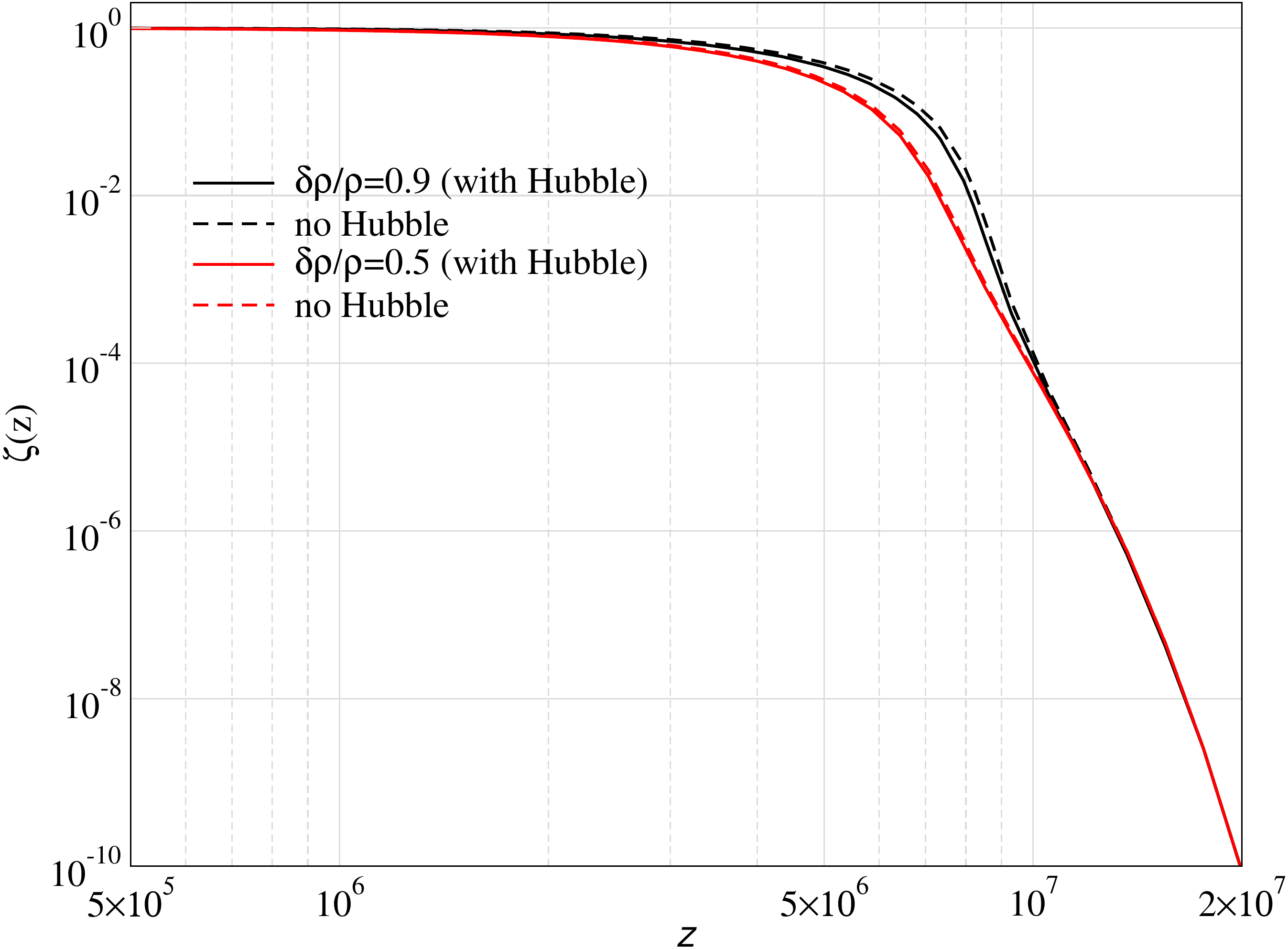}
\\
\caption{Comparison of the distortion visibility function for decaying particle as a function of lifetime with and without modification to the Hubble expansion for large energy release. }
\label{fig:visibility_decay_hubble}
\end{figure}


In this section, we study constraints on decaying particle energy injection scenarios. The energy injection rate is given by Eq.~\eqref{eq:decay_source}. We choose $\DlnrhoCMBt$ to fix the parameter $f_X$ for a given lifetime, which we set to $t_X=t(z_X)$ assuming the standard expansion history. We include the changes to Hubble parameter in the thermalization calculation as described in Sec.~\ref{subsec:initial_condition}.

In Fig.~\ref{fig:visibility_decay}, we present the distortion visibility for the decaying particle scenario as a function of $z_X$ for various values of $\DlnrhoCMBt$. 
For $z_X\lesssim 10^7$, the visibility shows a  behaviour that is qualitatively similar to the single injection cases. The departure from the small distortion solution is slightly less dramatic. 
This is because the energy release is extended over a significant redshift range, which lowers the typical distortion amplitude at fixed $\DlnrhoCMBt$ and hence reduces non-linear effects. 

The most dramatic difference between the single injection and the decay cases is the behaviour of the visibility function at $z\gtrsim 10^7$. There the visibility for the decaying scenario approaches the small distortion limit irrespective of $\DlnrhoCMBt$. 
This was not anticipated by CRA20, where it was speculated that the visibility corrections could be similar to the ones for single injection.
The explanation of this behaviour can be found in Fig.~\ref{fig:T_e_decay}, where we illustrate the electron temperature evolution. At $z\gtrsim 10^7$, DC and BR are extremely efficient in producing soft photons which erase the existing distortion. Therefore, even if we are constantly adding more and more energy to the existing CMB, there is never a significant buildup of distortion which will make the non-linear aspects  manifest. This can be seen for the cases $z_X=10^7$ and $2\times 10^7$ for which the maximal fractional increase in electron temperature is at the sub-percent level. But for $z_x\lesssim 10^7$, there is a buildup of distortion, as expected, which then leads to non-linear corrections. 

In Fig.~\ref{fig:constraints_decay}, we plot the constraints on $\DlnrhoCMBt$ for the decaying particle scenario from the non-linear Kompaneets treatment. Overall, the constraints do not depart much from the small distortion limit. For $z_X\gtrsim 10^7$, this was explained above. However, even for $z_X\lesssim 10^7$, there is practically no difference. This is because when $\zeta(z)\gtrsim 10^{-4}$, only $\DlnrhoCMBt\lesssim 0.6$ is allowed for \COBEF (for \PIXIE, it is a lot less). We thus only expect corrections to arise from a small region around 
$z_X\simeq \pot{6}{6}-10^7$, since only there the allowed $\DlnrhoCMBt\gtrsim 10^{-2}$. The differences are just about visible in Fig.~\ref{fig:constraints_decay} for \COBEF but the effect is negligible for \PIXIE.

We close our discussion by illustrating the effect of changes to the expansion rate from decaying particles. 
For this, we show the distortion visibility function with and without this modification in Fig.~\ref{fig:visibility_decay_hubble}. Even for large release, we do not see significant changes in the visibility. The main effect is just a change of the mapping between $t$ and $t_X$. Therefore, the constraints are expected to be similar even without modifying the Hubble rate from energy injection. However, given other uncertainties in the modeling of decaying particle distortions \citep[e.g.,][]{Kawasaki2005, Acharya2019a} we did not explore this aspect any further.

\begin{figure}
\centering 
\includegraphics[width=\columnwidth]{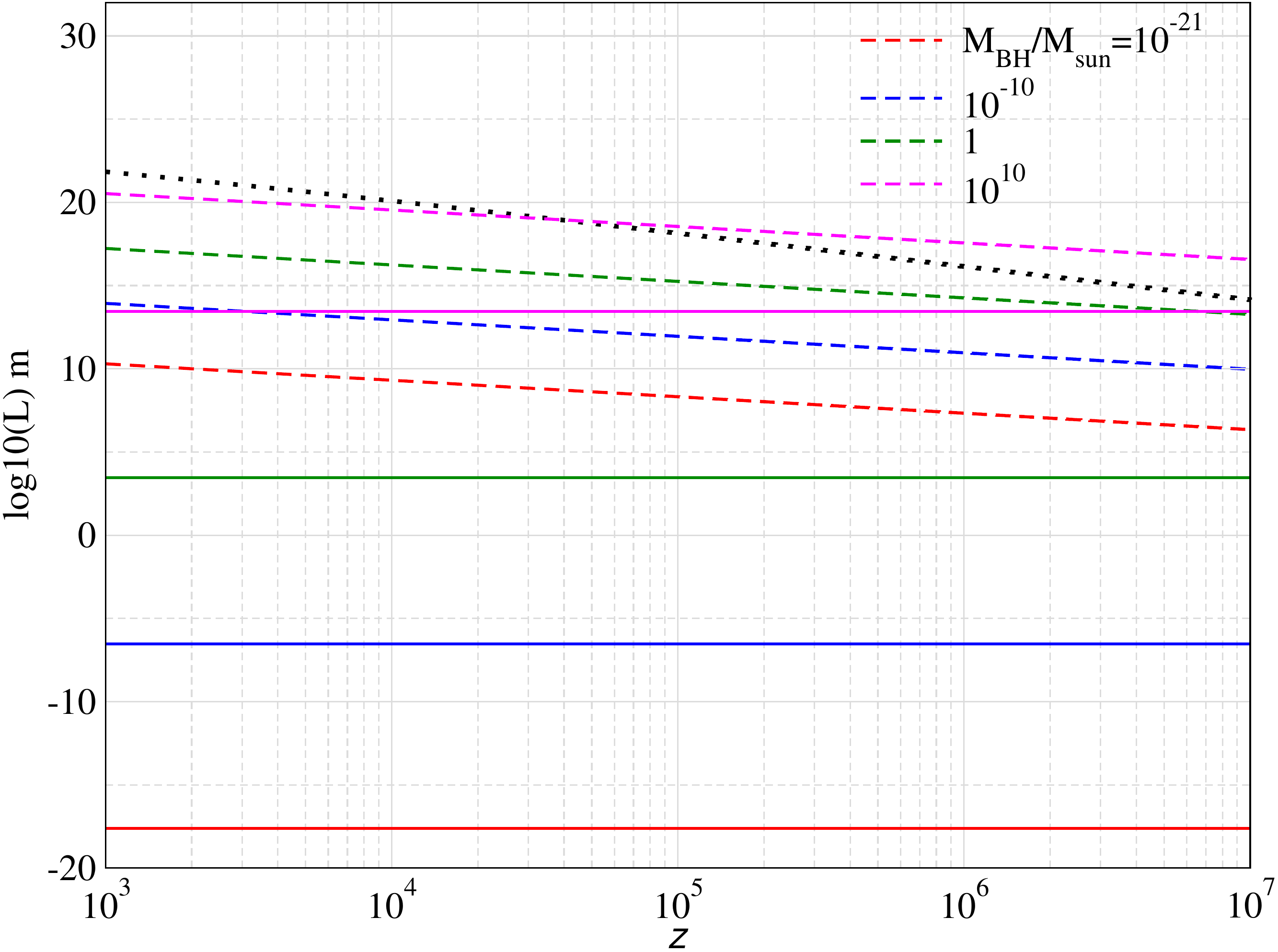}
\caption{Comparison of black hole spacing, $L_{\rm PBH}$ (dashed lines) with Schwarzschild radius, $R_{\rm S}$ (solid lines) and horizon size, $L_H=c/H(z)$ (dotted line). For all shown cases, $L_{\rm PBH}\gg R_{\rm S}$, implying that energy injection is highly localized. Whenever $L_H\gg L_{\rm PBH}$, many PBHs of the given mass are inside a Hubble patch.}
\label{fig:blackhole_spacing}
\end{figure}


\section{Importance of localization}
\label{sec:localization}
When carrying out the thermalization calculation, we usually assume that the energy is injected uniformly. However, if the injection happens in a localized and anisotropic manner, the maximal (local) distortion amplitude is underestimated and thus the large distortion regime might never me reached even if present. For PBHs and possibly high-energy particle cascades, this could be relevant.

Let us ask when localization effects become important for PBHs. For this we need to estimate the average distance between black holes and compare it with the relevant length scale on which the energy is injected. For an evaporating black hole, the emitted energetic photons can travel a long distance before depositing their energy in an ionized universe. In that case, the photon travel distance in a Hubble time is the relevant scale though we need to solve the radiative transfer problem to find the exact number. While soft photons emitted due to black hole super-radiance \citep{PL2013} can be trapped close to the horizon due to high optical depth to absorption by thermal electrons \citep{Chluba2015GreensII}.
Photon emission due to black hole accretion is more effective around recombination epoch \citep{AK2017}, because at high redshifts the radiation pressure blows out the infalling gas. But at these redshifts, we have neutral gas which can absorb the energetic photons and trap the radiation close to the horizon. 

Here, we give a brief estimate of the relevant length scales involved but defer a dedicated calculation for future.
Assuming the fractional abundance ($f_{\rm PBH})$ of black holes with mass ($M_{\rm PBH}$), the number density of black holes is given by, $N_{\rm PBH}=\frac{f_{\rm PBH}\rho_c(1+z)^3}{M_{\rm PBH}}$, The average distance between PBHs scales as $L_{\rm PBH}\simeq N_{\rm PBH}^{-1/3}$. The constraints on $f_{\rm PBH}$ from CMB spectral distortions are of the order of $10^{-4}-10^{-5}$ for evaporating \citep{Lucca2020,AK2020}, accreting black holes and from superradiance . For evaporating black holes, $\mu$-distortion constraints are relevant for a small mass range of $10^{11}-10^{13}$g ($\simeq 10^{-22}-10^{-20} M_{\odot} $). We also have constraints on a broader and heavier distribution ($M\gtrsim M_{\odot}$) of primordial black holes from accretion and superradiance.

Using these numbers, in Fig. \ref{fig:blackhole_spacing} we compare the average distance between the black holes with their Schwarzschild radius and the horizon scale as a function of redshift. For black holes with mass $10^{-21} M_{\odot}(\sim 10^{12}g)$ which evaporate at $z\simeq 10^5$, the horizon size is much larger than the Schwarzschild radius and the average distance between black holes, therefore, we expect the energy deposition to the CMB to be more homogeneous. But for $M\gtrsim M_{\odot}$, we see that localization effects may be important as the relevant physics for energy deposition to the CMB can be different. In this case, a treatment of large distortion effects may be highly relevant. In addition, spectral spatial effects are expected to become important.

\section{Large low frequency photon injection}

\label{sec:low_freq_photon}

\begin{figure}
\centering 
\includegraphics[width=\columnwidth]{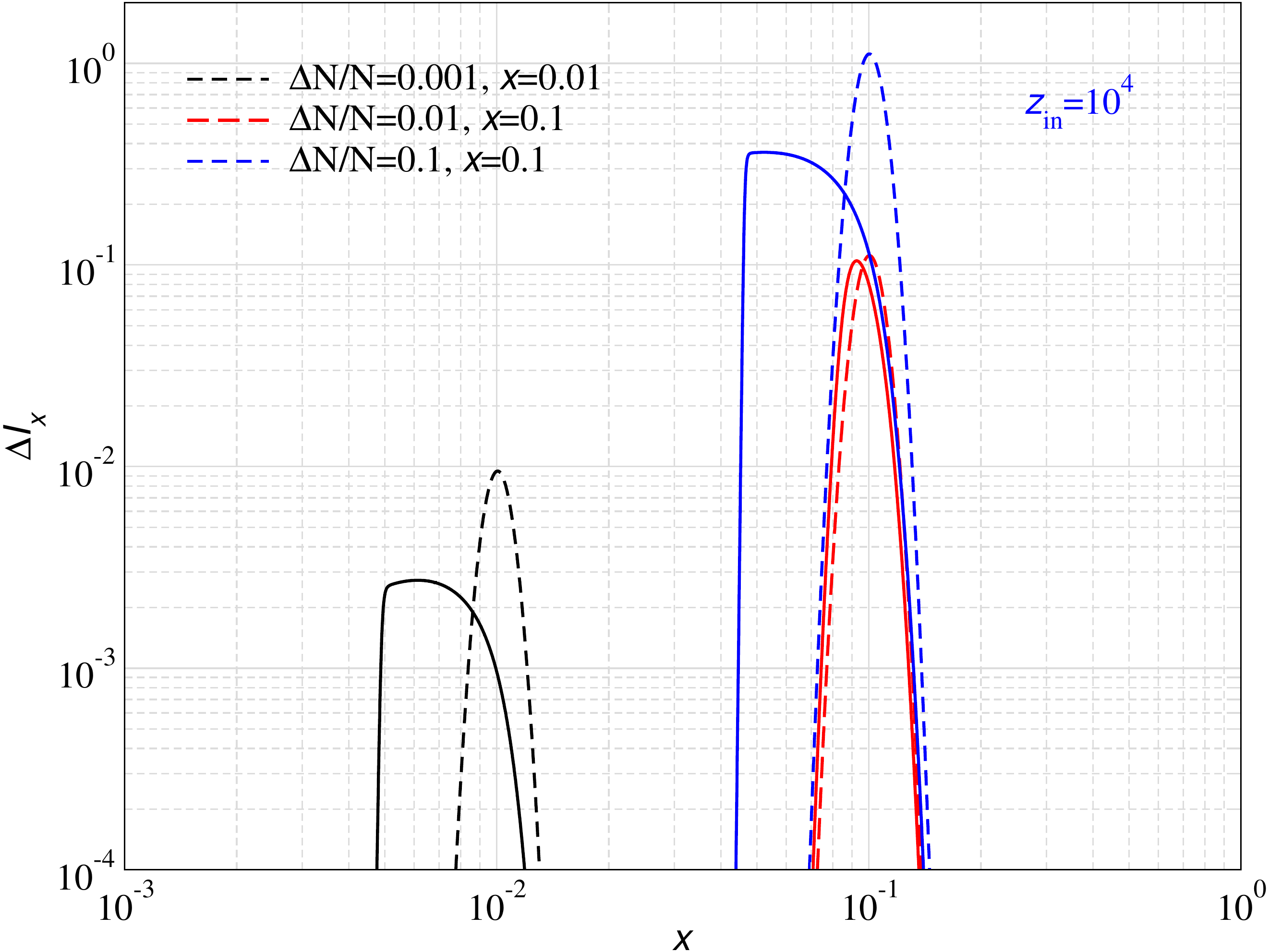}
\caption{Intensity of distortion at low frequencies for single photon injection at $z_{\rm in}=10^4$. The injected photon spectrum is a narrow Gaussian function ($1\%$ width) with mean at the denoted frequencies. Dashed and solid lines show linear and non-linear treatments respectively.}
\label{fig:spectrum_low_freq}
\end{figure}


\begin{figure}
\centering 
\includegraphics[width=\columnwidth]{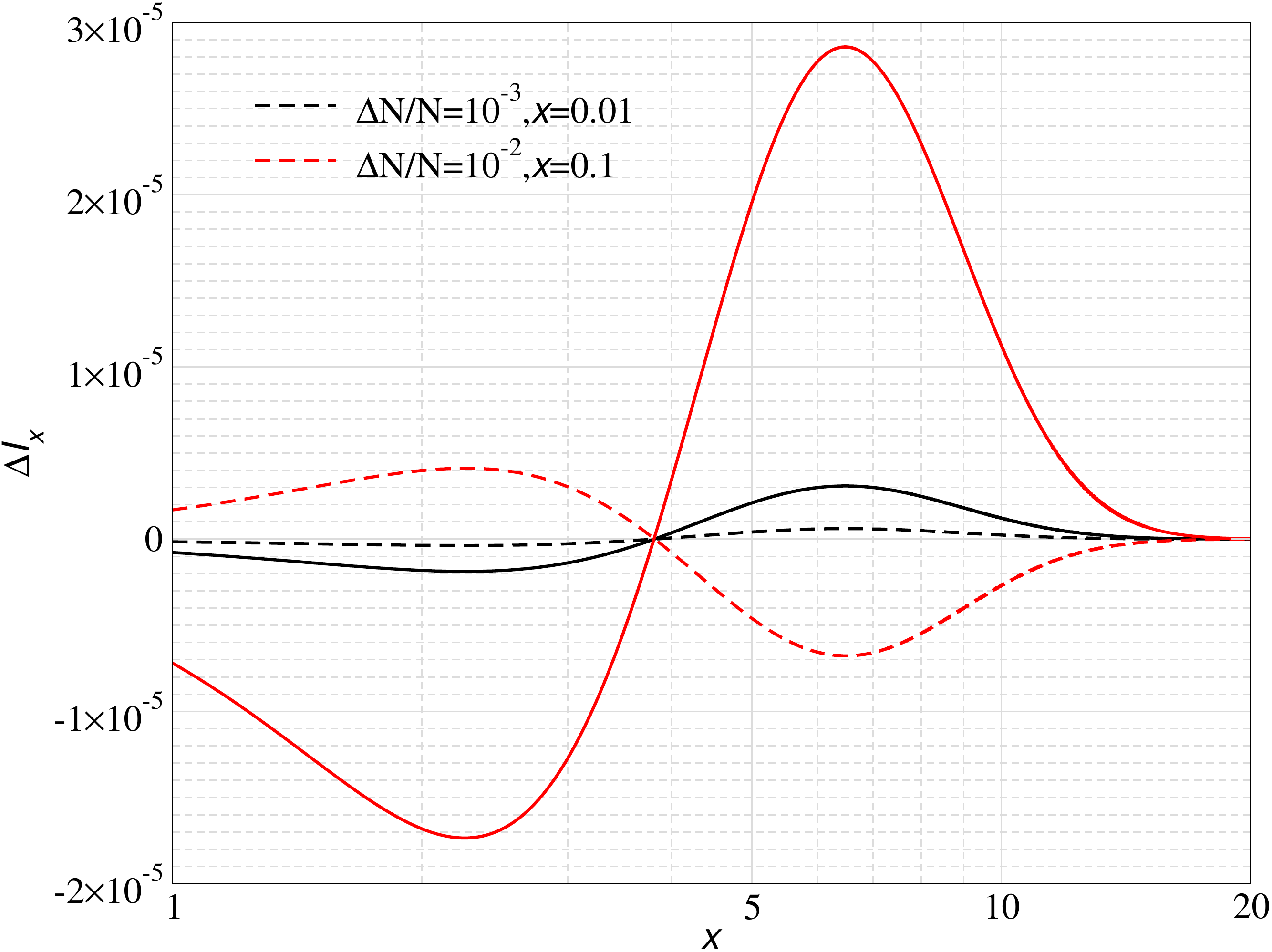}
\caption{Intensity of distortion at high frequencies for single photon injections as in Fig.~\ref{fig:spectrum_low_freq}. }
\label{fig:spectrum_high_freq}
\end{figure}


\begin{figure}
\centering 
\includegraphics[width=\columnwidth]{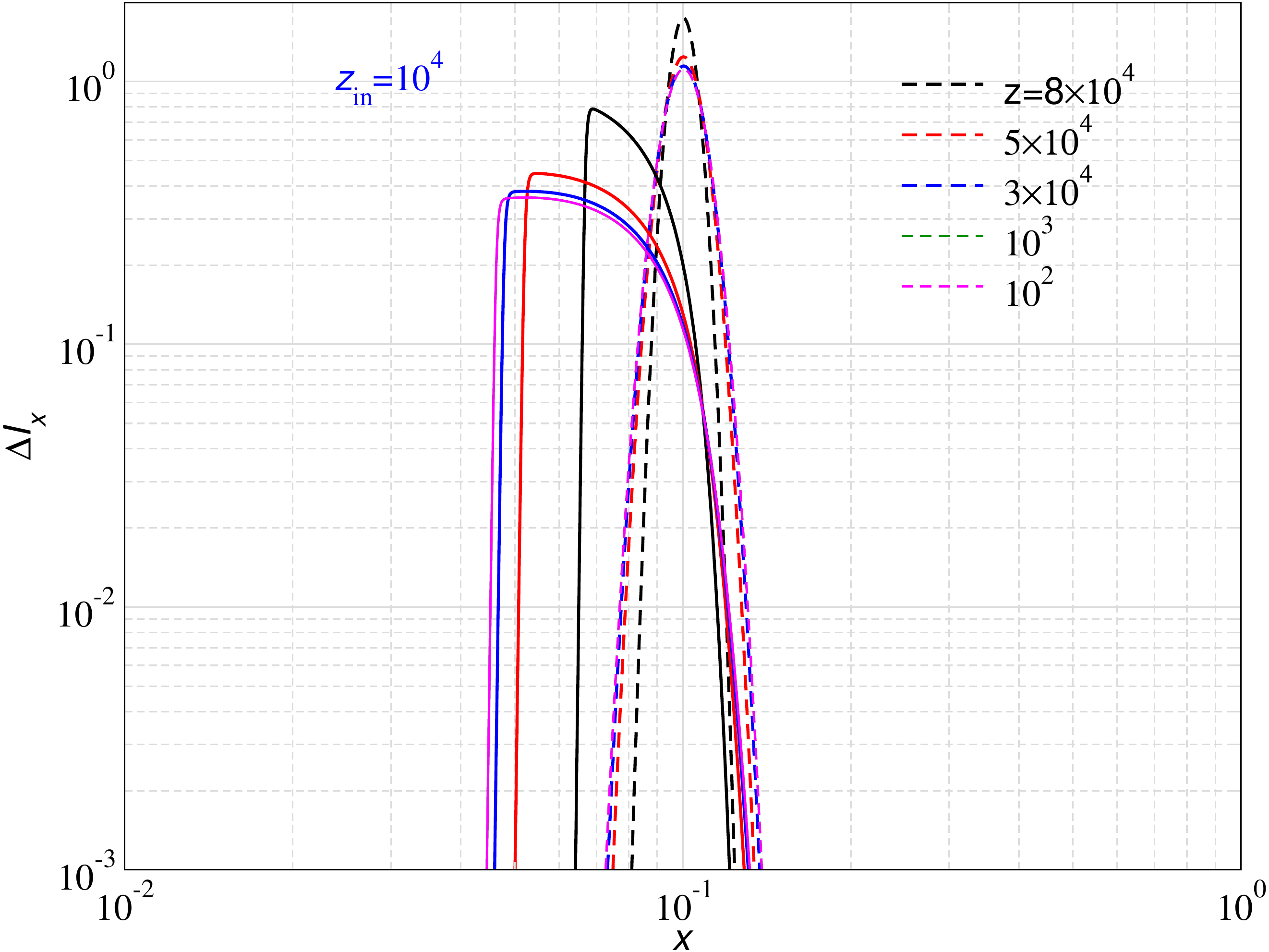}
\caption{Snapshot of evolution of photon spectrum as a function of redshift in linear approximation (dashed) and full non-linear calculations (solid). Photons are injected one time at $z=10^4$ at $x=0.1$ with $\Delta N/N=0.1$. }
\label{fig:spectrum_snapshot}
\end{figure}


In this section, we study a few cases of low-frequency photon injection and compare linear and non-linear solutions. This extends previous calculations \citep{Chluba2015GreensII, Bolliet2020PI} into the non-linear regime. Some additional discussion of non-linear photon injection cases can be found in \cite{Brahma2020}.

Some illustrative comparison is shown in Fig.~\ref{fig:spectrum_low_freq} and \ref{fig:spectrum_high_freq}. The choice of parameters in these figures are not entirely arbitrary. First, stimulated scattering effects are expected to become important at low frequencies, where a significant population of photons can principally be created without violating \COBEF constraints. 
In astrophysical situations, this can lead to interesting spectral shapes and modifies the evolution of the photon spectrum \citep{Sunyaev1970stimradio, Chluba2008d}.
We therefore focus on injection at $x\lesssim 0.1$.

Second, we are interested in distortions that are still visible today and only partially Comptonized. We thus consider injection at $z_{\rm in}=10^4$, where the effect of scattering is still noticeable but the distortion cannot be fully converted into a $\mu$- or $y$-type signal. 

Third, at $z_{\rm in}=10^4$, photons that are injected at $x\lesssim 10^{-3}-10^{-2}$, the photons are very likely to be absorbed by the background electrons via BR, raising their thermal energy and giving rise to $y$-distortion at high frequencies \citep{Chluba2015GreensII}. 
Even in these cases, one does expect changes to the time-scale on which the energy is reprocessed, since stimulated scattering terms \citep{Chluba2008d} will increase the photon absorption probability by more rapidly moving photons towards lower frequencies, however, we are not interested in these details here. We therefore consider injection frequencies $x_{\rm in}=10^{-2}$ and $0.1$ as instructive examples.

The non-linear effects manifest through the term $\propto n(n+1)$ of Eq.~\eqref{eq:Kompaneets_eq} with $n=n_{\rm pl}+\Delta n$. In the linear approximation, we drop the term proportional to $O(\Delta n^2)$ which unsurprisingly becomes important for $\Delta n/n_{\rm pl}\gtrsim 1$. 
Fixing the injected number of photons, implies an amplitude $\Delta n/n\propto \Delta N/N\,x_{\rm in}^{-2}$. For $\Delta N/N\simeq 0.001$, we therefore expect non-linear terms to be come important for $x_{\rm in}=10^{-2}$, while with $x_{\rm in}=0.1$ they should remain small even for $\Delta N/N=10^{-2}$
. Indeed we confirm this expectation in Fig.~\ref{fig:spectrum_low_freq}.
Naively, $\Delta N/N\simeq 10^{-3}$ and even $\simeq 10^{-2}$, may be believed to be  relatively small; however, this only parameterizes the relative number density of photon injection w.r.t to the total number density of the CMB. Since the injected photon spectrum is a sharply peaked function, $\Delta n/n_{\rm pl}$ can still be much greater than 1, locally at a particular frequency $x$.
By increasing to $\Delta N/N=10^{-1}$ at $x_{\rm in}=0.1$, we again find the non-linear corrections to become important.

Another important feature of the non-linear treatment is the shock-like shape on the low-frequency side of the distortion (see Fig.~\ref{fig:spectrum_low_freq}). Photons pile-up due to non-linear effects as they on average are moving downward towards lower frequencies. This shock-like behavior is completely absent in the linear treatment. To numerically resolve the 'shock' we had to increase the total number of frequency points to $\simeq 16,000$ between $x_{\rm min}=10^{-4}$ and $x_{\rm max}=100$. Improvements in the numerical treatment in particular for the PDE approach might be needed, but we leave a more detailed exploration to future work.

The non-linear stimulated term can move the injected photons to significantly lower energy compared to the linear case. The extra energy is gained by the background electrons which shows up as positive $y$-distortion. For the examples shown in Fig.~\ref{fig:spectrum_low_freq}, in the linear treatment the electrons on average cool as their average energy is higher than the injected photon energy. 
For the combination of parameters $x=0.01, \Delta N/N=10^{-3}$ and $x=0.1, \Delta N/N=0.01$, the high-frequency $y$-distortion in the linear case has an amplitude $\Delta I_x\simeq 10^{-6}-10^{-5}$ (see Fig.~\ref{fig:spectrum_high_freq}), while for $x=0.1, \Delta N/N=0.1$, it is $\simeq 10^{-3}$. The latter combination is indeed already excluded by \COBEF, and was therefore not shown here. The linear case with $x=0.01, \Delta N/N=10^{-3}$ has a  {\it positive} $y$-distortion while for $x=0.1$, we find a {\it negative} $y$-distortion. This is because the photons are likely to be absorbed at $x\lesssim 10^{-2}$ which give rise to {\it positive} $y$-distortion, as explained above.

Switching to the non-linear treatment, for $x=0.1, \Delta N/N=0.01$ we now find that the high-frequency $y$-type distortion switches sign (see Fig.~\ref{fig:spectrum_high_freq}). This indicates that photons actually down-scatter more efficiently leading to a heating of the electrons. Indeed, the low-frequency distortion shape supports this conclusion. 
For the case for $x=0.01, \Delta N/N=0.001$, we notices an amplification of the high-frequency distortion amplitude. This is because more photons can be absorbed by BR given that stimulated scattering on average moves the photon distribution towards lower frequencies, where the absorption probability is larger.

Finally, in Fig.~\ref{fig:spectrum_snapshot}, we show snapshots for the evolution of the solution for $x=0.1, \Delta N/N=0.1$ at few different times. The shock-like structure quickly develops, leading to extra broadening of the distortion in comparison with the linear treatment. 

A comprehensive study on the constraints from photon injection scenarios was done in \cite{Bolliet2020PI}; however, the authors assumed the distortions to be linear. 
Our results show that non-linear terms can qualitatively change the resultant distortion at high frequencies and also the specific distortion shape and dynamics at low-frequencies. This is expected to modified the corresponding distortion limits, motivating a more detailed calculation.
%

\section{Conclusions}
\label{sec:conclusion}
In this paper, we computed the CMB spectral distortion solutions from large energy injections, going beyond the small distortions approximation which is typically used in the literature. For this, we developed two independent numerical schemes for {\tt CosmoTherm}. In the first, we solve the full Kompaneets equation with the non-linear terms. The Kompaneets equation assumes Compton scattering to be non-relativistic which may fail for large energy injections at high redshifts. In the second, we use the exact Compton scattering kernel from {\tt CSpack}, we circumvent this limitation. We also explicitly treat the time-evolution of photon spectrum and electron temperature, both aspects that were not accounted for in CRA20. 

We study two energy injection scenarios: single energy injections and decaying particle cases. To derive constraints, we compute the distortion visibility function. Due to the non-linear nature of solutions, the visibility is a function of amount of energy injected to the CMB as opposed to small distortions approximation (Fig. \ref{fig:visibility_delta}). There is a higher probability for distortions to survive for larger energy injections due to the complicated interplay between Compton scattering and the photon emission/absorption processes (see also CRA20).
Consequently, the constraints on single injection cases are strengthened compared to the small distortion approximation (Fig.~ \ref{fig:constraints_delta}). 

For illustration, we also briefly consider the evolution of distortions for large energy extraction (see Sect.~\ref{sec:single_energy_extraction}). This could be relevant to distortion constraints from dark matter interactions \citep{Yacine2015DM, Yacine2021}. Due to the cooling of photons, strong non-linear effects become important (see Fig.~\ref{fig:snapshot_Drho}). Due to these effects, the distortion visibility function for large energy extraction is expected to differ from the one for energy injection. However, a more detailed computation is left to future work.

Somewhat surprisingly, for decaying particles we obtain constraints similar to the small distortion approximation (Fig. \ref{fig:constraints_decay}). This is due to the dilution of energy injection over a broad redshift range and the  photon non-conserving process becoming very efficient at $z\gtrsim 10^7$. 
Similarly, we expect that other energy injection scenarios which have a broad redshift distribution can be treated using the small distortion approximation. One example is the dissipation of acoustic modes from large power spectrum enhancements \citep{Chluba2012inflaton, Byrnes2019}; however, case-by-case calculations are required to obtain accurate results. Relativistic corrections to Compton scattering can become important for single injection cases while they remain subdominant for the decay case.

Several physically-motivated scenarios, which can be thought of as single energy injection, exist. For those, modifications of the distortion limits due to large distortion effects are expected.
Examples are black hole evaporation \citep{H1975} and photon injection from black hole super-radiance \citep{Zeldovich1971,TP1974}. Even though, theoretically, black holes evaporate over a broad redshift range, most of the emission happens as a burst over a very short redshift range \citep{Poulin2017, Lucca2020,AK2020}. 
Similarly, the energy extraction from the black hole due to super-radiance can be of extremely fast at $z\gtrsim 10^3$ \citep{PL2013,BW2020}. 

Energy injection processes are furthermore highly localized near black holes (see Sect.~\ref{sec:localization}). One typically assumes that all the injected energy is distributed uniformly over the cosmological volume. But since the injection process can be highly localized, this approximation dilutes the injected energy density or, in other words, can 'linearize' the distortions. 
Our modelling is thus one step towards a proper treatment for such localized bursts of large energy injection.

We also performed a brief qualitative study for monochromatic low-frequency photon injections at $z\simeq 10^4$ (see Sec. \ref{sec:low_freq_photon}). This demonstrates that non-linear effects can be significant for late photon injection  without violating current constraints \citep{Fixsen1996}. We obtain qualitatively different solutions compared to the small distortion approximation in the frequency range $0.01\lesssim x\lesssim 10$ (Fig.~\ref{fig:spectrum_low_freq}). The low-frequency solution is expected to be important for 21 cm calculations from low-frequency photon injections. This could be particularly relevant given the possible presence of the ARCADE radio excess \citep{Fixsen2011excess} and its link to the EDGES observation \citep{Chang2018}.
We hope to carry out a detailed study without assuming linearity of the problem in future.

\vspace{-3mm}
{\small
\section*{Acknowledgments}

This work was supported by the ERC Consolidator Grant {\it CMBSPEC} (No.~725456).
JC was furthermore supported by the Royal Society as a Royal Society University Research Fellow at the University of Manchester, UK (No.~URF/R/191023).
}

\section{Data availability}
The data underlying in this article are available in this article.

{\small
\vspace{-3mm}
\bibliographystyle{mn2e}
\bibliography{Lit}
}
\newpage

\end{document}

%% file: Befehle.tex
\newcommand{\GHz}{{\rm GHz}}

\newcommand{\expf}[1]{{{\rm e}^{#1}}}

\newcommand{\eV}{{\rm eV}}
\newcommand{\cm}{{\rm cm}}

\newcommand{\Tz}{{T_{z}}}

\newcommand{\TCMB}{T_{\rm CMB}}

\newcommand{\nbb}{n^{\rm pl}}

\newcommand{\xe}{x_{\rm e}}

\newcommand{\id}{{\,\rm d}}

\newcommand{\beq}{\begin{equation}}   %

\newcommand{\eeq}{\end{equation}}   %

\newcommand{\beqa}{\begin{eqnarray}}   %

\newcommand{\eeqa}{\end{eqnarray}}   %

\newcommand{\beal}{\begin{align}}

\newcommand{\enal}{\end{align}}

\newcommand{\bspl}{\begin{split}}

\newcommand{\espl}{\end{split}}

\newcommand{\bsub}{\begin{subequations}}

\newcommand{\esub}{\end{subequations}}

\newcommand{\bmulti}{\begin{multline}}   %

\newcommand{\beqm}{\begin{mathletters}}   %

\newcommand{\eeqm}{\end{mathletters}}   %

\newcommand{\kB}{k_{\rm B}}

\newcommand{\me}{m_{\rm e}}

\newcommand{\Ne}{N_{\rm e}}

\newcommand{\Te}{T_{\rm e}}

\newcommand{\The}{\theta_{\rm e}}

\newcommand{\sigT}{\sigma_{\rm T}}

\newcommand{\pot}[2]{#1 \times 10^{#2}}

\newcommand{\Thz}{\theta_{z}}
